\documentclass{aa}

\usepackage{newtxtext,newtxmath}
\usepackage[T1]{fontenc}
\usepackage{ae,aecompl}

\usepackage{natbib}
\bibpunct{(}{)}{;}{a}{}{,} % to follow the A&A style

\usepackage{graphicx}	% Including figure files
\usepackage{amsmath}	% Advanced maths commands
\usepackage{amssymb}	% Extra maths symbols
\usepackage{xcolor}
\usepackage[normalem]{ulem}
\usepackage{subcaption}
\usepackage{booktabs}

\providecommand{\gaia}{\textit{Gaia }}

\providecommand{\kms}{km s$^{-1}$ }
\providecommand{\degr}{$^{\circ}$ }
\providecommand{\msun}{$M_\odot$}

\newcommand{\kmskpc}{km s$^{-1}$ kpc$^{-1}$}
\newcommand{\vlos}{$V_{\rm los}$}
\newcommand{\los}{\mbox{l.-o.-s.}}
\newcommand{\omp}{$\Omega_p$}
\newcommand{\vp}{$V_\phi$}

\begin{document}

\title{The bar pattern speed of the Large Magellanic Cloud}

\author{Ó. Jiménez-Arranz\inst{1,2,3}
   \and L. Chemin\inst{4}
   \and M. Romero-Gómez\inst{1,2,3}
   \and X. Luri\inst{1,2,3}
   \and \\ P. Adamczyk\inst{5}
   \and A. Castro-Ginard\inst{6}
   \and S. Roca-Fàbrega\inst{7,8}
   \and P. J. McMillan\inst{7}
   \and M.-R. L. Cioni\inst{9}
}

\institute{{Departament de Física Quàntica i Astrofísica (FQA), Universitat de Barcelona (UB), C Martí i Franquès, 1, 08028 Barcelona, Spain}
\and
{Institut de Ciències del Cosmos (ICCUB), Universitat de Barcelona, Martí i Franquès 1, 08028 Barcelona, Spain}
\and
{Institut d’Estudis Espacials de Catalunya (IEEC), C Gran Capità, 2-4, 08034 Barcelona, Spain}
\and
{Instituto de Astrofísica, Universidad Andres Bello, Fernandez Concha 700, Las Condes, Santiago RM, Chile}
\and
{Centro de Astronomía - CITEVA, Universidad de Antofagasta, Avenida Angamos 601, Antofagasta 1270300, Chile}
\and
{Leiden Observatory, Leiden University, Niels Bohrweg 2, 2333 CA Leiden, The Netherlands}
\and
{Lund Observatory, Division of Astrophysics, Department of Physics, Lund University, Box 43, SE-221 00 Lund, Sweden}
\and
{Departamento de Física de la Tierra y Astrofísica, UCM, and IPARCOS, Facultad de Ciencias Físicas, Plaza Ciencias, 1, Madrid, E-28040, Spain}
\and
{Leibniz-Institut für Astrophysik Potsdam, An der Sternwarte 16, D-14482 Potsdam, Germany}
}

\date{Received <date> / Accepted <date>}

\abstract 
{The Large Magellanic Cloud (LMC) internal kinematics have been studied in unprecedented depth thanks to the excellent quality of the \gaia mission data, revealing the disc's non-axisymmetric structure.}
{We want to constrain the LMC bar pattern speed using the astrometric and spectroscopic data from the \gaia mission.}
{We apply three methods to evaluate the bar pattern speed: it is measured through the Tremaine-Weinberg (TW) method, the Dehnen method and a bisymmetric velocity (BV) model. The methods provide additional information on the bar properties such as the corotation radius and the bar length and strength. The validity of the methods is tested with numerical simulations.}
{A wide range of pattern speeds are inferred by the TW method, owing to a strong dependency on the orientation of the galaxy frame and the viewing angle of the bar perturbation. The simulated bar pattern speeds (corotation radii, respectively) are well recovered   by the Dehnen method (BV model). Applied to the LMC data, the Dehnen method finds  a pattern speed $\Omega_p = -1.0 \pm 0.5$ \kmskpc, thus corresponding to a bar  which barely rotates, slightly counter-rotating with respect to the LMC disc. The BV method finds a LMC bar corotation radius of $R_c = 4.20 \pm 0.25$ kpc, corresponding to a pattern speed $\Omega_p = 18.5_{-1.1}^{+1.2}$ \kmskpc.}
{It is not possible to decide which global value best represents an LMC bar pattern speed with the TW method, due to the strong variation with the orientation of the reference frame. The non-rotating bar from the Dehnen method would be at odds with the structure and kinematics of the LMC disc. The BV method result is consistent with previous estimates and gives a  bar corotation-to-length ratio of  $1.8 \pm 0.1$, which makes the LMC hosting a slow bar.}


%\begin{document}

\maketitle

\section{Introduction}

The angular speed of stellar bars is a fundamental parameter in dynamics of galaxies from which the principal bar-disc resonances can be identified, and the structure of stellar orbits can be studied in a given gravitational potential \citep[e.g.][]{1989contopoulos,2008binney}. Furthermore, the bar pattern speed is thought to reflect certain properties of the halo density at low radius because bars and haloes of stars and/or dark matter are believed to continuously interact during galaxy evolution through dynamical friction. Numerical models and observations suggest that the bar speed is slower when the inner halo is denser \citep{2000debattista, 2023buttitta}. The pattern speed of simulated bars is  also seen to slow down with the secular evolution, as opposed to its growing length and strength \citep{2014sellwood}. 

The importance of measuring bar pattern speeds in the dynamics and evolution of galaxies has thus grown significantly with the emergence of large-scale long-slit and 3D spectroscopic surveys, which have enabled the estimations of a few hundreds of pattern speeds \citep[][and references therein]{2019cuomo, 2019guo,2023geron}. These works found that most stellar bars are fast, i.e. bars show a ratio of corotation-to-bar radius smaller than 1.4  \citep[following][]{Athanassoula1992}, and that high angular speeds are for small and weak bars, and within faint galaxies \citep{2020cuomo}. However, the large occurrence of fast bars is a critical issue for the simulations made in a cosmological context, as the inner density of simulated dark matter halos is so cuspy that the parent discs should host mostly slow bars. This discrepancy suggests a failure in  cosmological simulations \citep{2021roshan}, or, perhaps, a bias in estimating pattern speeds  of bars from observations \citep{fragkoudi21}. This highlights the difficult part of measuring pattern speeds from observation since it represents a single time instant in the entire evolution of galaxies.

Bar pattern speeds from these studies have been inferred  exclusively by means of the  Tremaine-Weinberg method \citep[][hereafter TW]{tw84}. Its direct application makes use of integrals of kinematics and positions of a tracer that should obey the continuity equation, like stars. Alternatively, \citet{dehnen23} proposed a new method for determining bar pattern speeds in numerical simulations using a single time snapshot. This involves measuring the Fourier amplitudes of particle positions and velocities within the bar region. In another work, \citet{drimmel22} proposed an indirect measurement of the bar angular speed by fitting a  bisymmetric model to the tangential velocities to get the bar phase angle   and corotation radius. Unfortunately, these last two methods have limited applicability due to their reliance on the availability of objects with individually measured planar (tangential and radial) velocities, which are rare. Only the Milky Way (MW) and the Large Magellanic Cloud (LMC) offer this possibility. To our knowledge, the \citet{dehnen23} formalism has never been applied to observations, while the method in \citet{drimmel22} estimated the bar corotation, orientation and pattern speed for the MW, using data from the \gaia mission \citep{2016prusti, gaiadr3}. Interestingly, they found a  Galactic bar speed consistent with the value inferred with the TW method \citep{Bovy2019}.

Our objective in this study is to determine the pattern speed and corotation region of the stellar bar in the LMC. The LMC is a dwarf spiral (or irregular) galaxy and one of the closest and brightest satellites of the MW, in interaction with the Small Magellanic Cloud (SMC) and the MW. It could thus be a challenge to study the  properties of a system like the LMC, because of its structure and velocity field impacted by the interaction, at least in the outermost regions \citep[e.g.][]{2019belokurov,luri20}. It is however a unique object  of its kind  for testing the three aforementioned methods due to the availability of different kinds of kinematic data for the entire disc. To achieve the objective, we are taking advantage of the opportunity to use both astrometric and spectroscopic  data from the \gaia mission. First, the TW method can be applied to line-of-sight (\los) velocities of thousands of LMC stars, as measured by the \gaia Radial Velocity Spectrometer \citep{Katz2022,jimenez-arranz23}, yielding \emph{a single value} of the LMC bar pattern speed, similarly to other galaxies. Second, the \gaia astrometric data  allow us to estimate  the two components  of the velocity tensor for millions of stars in the LMC plane  \citep{luri20}. Unlike the \los\ velocities given in the sky frame,  these  two components make it possible to apply the Dehnen method, the bisymmetric model, as well as a modified version of the TW method, adapted to the disc plane. In this latter case, \emph{multiple values} of the LMC bar pattern speeds can be inferred, corresponding to multiple orientations of the Cartesian frame of the LMC plane, thus multiple viewing angles of the bar in the disc reference frame. This allows us to  compare various estimates of the bar pattern speed, and assess the validity of the methods.

The paper is organised as follows. In Section \ref{sec:methods}, we describe the methods used to measure the bar pattern speed. In Section~\ref{sec:simulations}, we validate the methods using two N-body simulations, one representing an isolated disc and the other an interacting disc. In Section~\ref{sec:results}, we apply the three methods to the LMC sample \citep{jimenez-arranz23}, to try to determine the pattern speed of the LMC bar. In Section~\ref{sec:discussion}, we discuss the implications of our findings for our understanding of the LMC and barred spiral galaxies in general. Finally, in Section~\ref{sec:conclusions}, we summarise the main conclusions of this work.

\section{Methods}
\label{sec:methods}

In this section, we describe the three methods applied to the \gaia data to infer the bar pattern speed of the LMC. The first  is the Dehnen method (Sect.~\ref{subsec:method_dehnen}), which can be applied to astrometric data. Secondly, we describe the Tremaine-Weinberg method (TW, Sect.~\ref{subsec:method_tw}), which can be applied to astrometric or spectroscopic data. Finally, we describe the bisymmetric velocity method (BV, Sect.~\ref{subsec:method_fourier}), which can only be applied to astrometric data.

In the Cartesian frame of the galaxy, the methods assume that the disc is in equilibrium (which may not be fully the case for the LMC), rotation is done around the $z$-axis, the kinematic center is located at the origin of the Cartesian coordinate frame, and the density is stationary in the frame rotating at \omp.  Furthermore, it is also assumed that the region where \omp\ is constrained should only contain the bar, that  is well distinguishable from other structures in the galaxy, such as spiral arms. It is worth mentioning that other density perturbations may exist in the region where the bar influences the stellar dynamics, and thus could impact the estimation of the bar \omp.

\subsection{The Dehnen method}
\label{subsec:method_dehnen}

The first method we use in this work is that of \citet{dehnen23}. We use the version of the code made publicly available with the paper. Here we summarise some of the main aspects of the method.

\citet{dehnen23} developed an unbiased, precise, and consistent method that simultaneously measures \omp\ and the orientation angle $\phi_b$ of the bar from single snapshots of simulated barred galaxies. These parameters are found assuming that the continuity equation applies: 
\begin{equation}\label{eq:tw2d}
    \frac{\partial \Sigma v_x}{\partial x} + \frac{\partial \Sigma v_y}{\partial y} + \frac{\partial \Sigma}{\partial t} = 0
\end{equation}
where $v_x$ and $v_y$ are the disc's velocity components in Cartesian coordinates $(x,y)$, where $\Sigma = \Sigma(x,y,t) = \Sigma(R,\phi-\Omega_p t)$ is the disc surface density,  ($R,\phi$) the corresponding cylindrical coordinates, and  $\Omega_p$  is the  angular speed of the rotating frame of the bar perturbation,  considered invariant with time. The method assumes that the centre of rotation is known, that the rotation is around the z-axis and that the density is stationary in the rotating frame. With these assumptions, $\partial \Sigma / \partial t = -\Omega_p \partial \Sigma / \partial \phi$ and Eq.~\ref{eq:tw2d} becomes :
\begin{equation}
\label{eq:tw2d_v2}
    \Omega_p \frac{\partial \Sigma}{\partial \phi} = \frac{\partial \Sigma v_x}{\partial x} + \frac{\partial \Sigma v_y}{\partial y} \, .
\end{equation}

This expression is the traditionally used in the TW method (see Section~\ref{subsec:method_tw} below). Here we use the Fourier method, as implemented in the public code, which consists of multiplying Eq.~\ref{eq:tw2d_v2} by the weight function $w(x)=W(R)e^{-im\phi}$, where $m$ is the azimuthal wave number and $W(R)$ a smooth window function (see their Eq.~25), and integrating over all space. The smooth window function is necessary to avoid issues at the edges of radial bins. The resulting expression for the pattern speed of an N-body model is the real value form (see details in their Appendix A) of the following equation:
\begin{equation}\label{eq:dehnen}
\Omega_p+\frac{i}{m}\frac{\dot{\Sigma}_m}{\Sigma_m} = 
\frac{\Sigma_i\mu_i\left[\dot{\phi}_iW_i+\frac{i}{m}\dot{R}_i(\partial W/\partial R)_i\right]e^{-im\phi_i}}{\Sigma_i\mu_iW_ie^{-im\phi_i}},
\end{equation}
where $\mu_i$ are the individual particle masses.

In fact, the method is divided in two steps. First, it defines which particles belong to the bar region,  $[R_0, R_1]$, defined as a continuous range of radial bins with large  amplitude of the bisymmetric density perturbation of second order, and having a roughly constant phase angle  (see their Appendix B for details). Hereafter we will  refer to  $R_1$ as the bar radius or length, as it agrees well with the definition of best estimates for bar lengths in numerical simulations \citep{2023ghosh}. And second, once the bar region is determined, it computes the bar pattern speed and the bar phase angle together with their uncertainties using the covariance matrix of the real part of Eq.~\ref{eq:dehnen} with $m=2$, this is their Eq.~A4.

\citet{dehnen23} applied their method to a suite of N-body models of isolated barred spiral galaxies. By comparing the results to \omp\ calculated using time-centred finite-differences from three consecutive snapshots, they found that their method is reliable and accurate, provided that the bar region is well-determined and a smooth window function is utilised. 

\subsection{The Tremaine-Weinberg method}
\label{subsec:method_tw}

The second method we use is the Tremaine-Weinberg (TW) method \citep{tw84}. As mentioned above, the main assumption of the method developed is that the density and  kinematics of the tracer obey the continuity equation (Eq.~\ref{eq:tw2d}). The method is designed for galactic systems in equilibrium and with a single pattern (see its application to a simulated disc with a bar perturbation in App.~\ref{sec:app-tpsim}).
The best kinematic tracer satisfying this condition is the old stellar population of galaxies. Estimates of \omp\ of galaxies using the TW method were thus mostly obtained from absorption lines of stellar populations, as observed by means of optical long-slit or integral field spectroscopy \citep[e.g.][]{merrifieldkuijken95,1999gerssen,2003aguerri, 2004debattista,2015aguerri,2019cuomo}, although emission lines of interstellar gas for a few galaxies were used as well  \citep[e.g.][]{1999bureau, 2005hernandez, 2004rand, 2008meidt-b, 2009chemin, 2021williams}. 

As shown in \citet{luri20} and \citet{jimenez-arranz23}, we can select millions of stars in the LMC disc with proper motions, from which the in-plane components $v_x$ and $v_y$ can be measured. Therefore, Eq.~\ref{eq:tw2d_v2} can be solved directly for the LMC, unlike any other galaxies. Integrating it with respect to $x$ yields:
\begin{equation}
\label{eq:tw_vxvy}
\begin{split}
    \Omega_p = \frac{\langle  v_y  \rangle }{\langle x \rangle }, \hspace{0.5cm} \text{where} \hspace{0.1cm} & \langle v_y \rangle  = \frac{\int_{-\infty}^{+\infty} v_y (x,y) \Sigma (x,y) dx}{\int_{-\infty}^{+\infty} \Sigma (x,y) dx}, \\
    & \langle x \rangle  = \frac{\int_{-\infty}^{+\infty} x \Sigma (x,y) dx}{\int_{-\infty}^{+\infty} \Sigma (x,y) dx} \, .
\end{split}
\end{equation}

These integrals can be numerically solved by discretising the space, i.e., by summing the surface density and kinematics along $x-$wedges at different $y$ positions (hereafter, pseudo-slit). Then, the pattern speed $\Omega_p$ can be determined by doing a linear fit of $\langle  v_y  \rangle$ vs $\langle x \rangle$. Interestingly, a permutation of $x$ and $y$ can be done in Eq.~\ref{eq:tw_vxvy}, so that \omp\  can also be estimated from $\langle  v_x  \rangle$ vs $\langle y \rangle$. To keep the analysis simple, we defer to another study the test of this alternative derivation. 

The continuity assumption is independent on the choice of the Cartesian frame. This implies that  we can choose arbitrarily the orientation of the reference $x-y$  plane by rotating it around the $z-$axis, and measure the TW integrals of Eq.~\ref{eq:tw_vxvy} at various orientations. Only astrometric data can make such analysis possible, unlike spectroscopic data.  This is thus a good opportunity for us to assess for the first time  the effect of the viewing angle of the bar in the disc plane on the TW integrals of Eq.~\ref{eq:tw_vxvy} (Sect.~\ref{sec:simulations}), and on the LMC bar pattern speed (Sect.~\ref{sec:results}).

To get an unbiased value of the bar angular speed,  we must restrict the linear regression by selecting exclusively the integrals from the bar region. This is defined as the points located out to the radius $R_1$ obtained with the Dehnen method (Sect.~\ref{subsec:method_dehnen}), and the best fit of the pattern speed \omp\ arises from selecting the integrals with $(\langle x \rangle^2 + y^2)^{1/2} < R_1$, thus avoiding the outer disc that mostly traces the spiral structure, which is expected to show a lower angular speed from  the bar \citep[see e.g.][with the example of the grand design spiral NGC 1068]{2006merrifield}. 

Additionally, the LMC is  the only galaxy for which both transverse and \los\ kinematics are available.  Because of such a lack of galaxies having observed planar kinematics, \citet{tw84} have historically adapted Eq.~\ref{eq:tw_vxvy} to work with sky plane coordinates $(X,Y) = (x, y \cos i)$ and the  \los\ velocity $V_{\rm los} = v_y \sin i + v_z \cos i$, where $i$ is the galaxy inclination, leading to: 
\begin{equation}
\label{eq:tw_vlos}
\begin{split}
    \Omega_p \sin i = \frac{\langle  V_{\rm los}  \rangle }{\langle  X  \rangle}, \hspace{0.2cm} \text{where} \hspace{0.1cm} & \langle V_{\rm los} \rangle  = \frac{\int_{-\infty}^{+\infty} V_{\rm los} (X,Y) \Sigma (X,Y) dX}{\int_{-\infty}^{+\infty} \Sigma (X,Y) dX} ,\\
    & \langle X \rangle  = \frac{\int_{-\infty}^{+\infty} X \Sigma (X,Y) dX}{\int_{-\infty}^{+\infty} \Sigma (X,Y) dX} \, .
\end{split}
\end{equation}
with $\langle V_{\rm los} \rangle $ and $\langle X \rangle$ being the intensity-weighted means of the \los\ velocity and position of the tracer, respectively. These integrals can numerically be solved by discretising the space, selecting the disc areas parallel to the disc major axis, yielding a value of $\langle X \rangle$ and $\langle V_{\rm los} \rangle$ for each $Y$. Unlike the previous case, we cannot vary the orientation of the reference frame here because the \los\ kinematics is firmly attached to the unique position angle of line of nodes (disc major axis).  Then, $\Omega_p \sin i$ is the result of the linear fit of $\langle  V_{\rm los}  \rangle$ vs $\langle X \rangle$.  Similarly to the planar velocities, only $\langle X \rangle$ and $\langle V_{\rm los} \rangle$ from the bar region must be considered, thus by selecting the TW points inside the sky region where $R_1$ is projected. 

For clarity, we hereafter refer to the version of the TW method involving the planar velocities as the In-Plane TW method (IPTW, Eq.~\ref{eq:tw_vxvy}), and the one using $V_{\rm los}$ data as the \los\ TW method (LTW, Eq.~\ref{eq:tw_vlos}).

\subsection{Bisymmetric model of the tangential velocity}
\label{subsec:method_fourier}

In \citet{drimmel22}, indirect measurements of the pattern speed \omp\ were performed by searching for the corotation radius $R_{c}$ within a simulated barred galaxy. Here, a second order asymmetry of the tangential velocity field $V_\phi$ was fitted. Variations at low radius of the phase angle $\phi_{2,\rm kin}$ of the bisymmetry  were then studied to locate $R_c$. Ignoring the  first order perturbation (lopsidedness), the Fourier decomposition $V_{\phi,\text{mod}}$ is given by:
\begin{equation}
\label{eq:modbisym}
    V_{\phi,\text{mod}} (R,\phi) = V_{0}(R) + V_2 (R) \cos (2 (\phi -\phi_{2,\rm kin} (R))),
\end{equation}
where $V_{0}$ and $V_2$, which only depend on the galactocentric radius $R$, are the rotation curve of the disc and the amplitude of the bisymmetric perturbation, respectively. Despite its empirical nature, this method is based on the principle that the bar pattern heavily influences, if not entirely governs, the structure of stellar orbits and the velocity field within $R_{c}$. In the ideal case of only a barred perturbation, with no spiral arms,   the periodic orbits inside corotation, called the $x_1$ family, are elongated with the bisymmetric perturbation (they are the back-bone of bars), while  beyond corotation, the orbits are elongated perpendicularly to the bar major axis   \citep{1980contopoulos, 1989contopoulos}. Inside corotation, the tangential velocity is maximum (minimum)  perpendicularly to (along) the direction of elongation of the orbits, and the opposite outside corotation.  We thus expect $\phi_{2,\rm kin}$ roughly constant within the bar,  then changing significantly its orientation near corotation, by an angle of $\sim 90\degr$. This variation can be even larger  in presence of a winding spiral structure beyond corotation which perturbs the orbits and kinematics as well. Furthermore, the possible existence of other resonances before corotation, like the ultra-harmonic one \citep{1996buta}, and the development of spiral arms before or near the bar ends complicate the shapes, orientations and kinematics of periodic orbits inside corotation. In these cases, we can also expect a variation  of $\phi_{2,\rm kin}$  between the bar ends and  the corotation radius. Once $R_{c}$ is determined near the location where $\phi_{2,\rm kin}$   changes significantly, we can infer $\Omega_p$ with the angular velocity curve, $\Omega(R) = V_{0}/R$, since stars move at the same speed as the bar at $R_c$, thus $\Omega_p = \Omega(R_{c})$. One should note that the harmonic decomposition is reminiscent of the bisymmetric flow model applied to \los\ kinematics of barred spirals \citep{2007spekkens}. 

\citet{drimmel22} applied their recipe to data from the Third Data Release of \gaia \citep[DR3,][]{gaiadr3,gaiadr3vallenari}. They evidenced a region of steep change of phase angle of the kinematic bisymmetry of $\sim 70\degr$ over a range of 2 kpc. A comparison with a test-particle simulation in which the \gaia errors model was propagated made them find the MW corotation  at the radius where $\phi_{2,\rm kin}$ is minimum, just after the location of the sharp transition of phase angle mentioned above.  \citet{drimmel22}  estimated a Galactic bar orientation with respect to the Galactic Center-Sun direction of $\sim 20\degr$, the Galactic bar corotation at  $R \sim 5.4$ kpc, and a pattern speed of $\sim 38$ \kmskpc, in good agreement with previous measurements involving a modified version of the TW method \citep{Bovy2019}.

\section{Testing the methods with simulations}
\label{sec:simulations}

In this section we use a snapshot of a simulation of a MW-mass galaxy, with no external perturbations, and a snapshot of a simulation of a LMC-like system interacting with a SMC-mass and MW-like systems to apply and validate the Dehnen method, the two variations of the  TW method, and the BV model. 

First, we use the B5 N-body simulation of an isolated barred galaxy from \citet{rocafabrega2013}, which consists of a live disc of $5$ million particles and a Toomre parameter of $Q=1.2$, and a live NFW halo. The disc to halo mass ratio is the appropriate so that the simulation develops a strong bar and two spiral arms which are transient in time. The snapshot we use has a counter-clockwise rotating bar  with a pattern speed of $\Omega_p = 21.5\pm 0.1$ \kmskpc\ determined as the average of finite-differences on the rate of change of the phase angle of the bar major axis in three consecutive snapshots over the radial range of the bar. The quoted uncertainty on the bar pattern speed refers to the standard deviation of the  pattern speeds derived from the three successive snapshots. The simulation time step is 16 Myr, representing 6\% of the bar period, which is appropriate to infer a robust bar pattern speed. The pattern speed places the bar corotation resonance at $R_c= 8.3\pm0.05$ kpc, computed as the radius at which the angular frequency curve, $\Omega(R)= V_{0}/R$, of the particles is equal to the bar pattern speed, with  $V_{0}$  given by the bisymmetric model (see Eq.~\ref{eq:modbisym}). It corresponds to a   fast bar with a rotation rate of $R_c/R_1 = 1.1^{+0.01}_{-0.01}$.

Second, we use one simulation of the KRATOS (Kinematic Reconstruction of the mAgellanic sysTem within the OCRE Scenario) suite, a comprehensive suite of 12 sets of pure N-body simulations of isolated or single-interacting galaxies, for a total of 30 models (Jim\'enez-Arranz et al., in preparation). The simulation we use in this paper models both an LMC-like and an SMC-mass system in the presence of a MW-mass system. We model the LMC-like system as a stellar exponential disc of 1.2M stars embedded in a live dark matter NFW halo. We consider a disc and DM halo with a mass of $5 \times 10^9$\msun\ and $1.8 \times 10^{11}$\msun, respectively, in agreement with observations as discussed in \citet[][and references therein]{lucchini22}. The disc's Toomre $Q$ parameter is 1.0, i.e. slightly gravitationally unstable. The SMC-mass system is modelled as a simple NFW halo. Both dark matter and stellar particles in the SMC-mass system are generated at once following the NFW profile with a total mass of $1.9 \times 10^{10}$\msun\ \citep[][and references therein]{lucchini22}. For the MW-mass system, we only model its DM content since we are not interested in its stellar component but only in its gravitational effects on the LMC-SMC-like system. The DM mass of the MW-mass system is considered to be $10^{12}$\msun\ \citep[][and references therein]{bobylev23}. In this work, we analyse a snapshot of the simulation taken just after the LMC-like system suffered a second close encounter with the SMC-mass system that generated an off-centred and out-of-equilibrium bar in the LMC-like system. The bar has a counter-clockwise rotation with a pattern speed of $\Omega_p = 17.2 \pm 1.6$ \kmskpc\ measured as the difference of the rate of change of the phase angle of the bar perturbation, using three consecutive snapshots with a time interval of 2 Myr. The pattern speed places the bar corotation resonance at $R_c= 3.6^{+0.8}_{-0.5}$ kpc. It corresponds to a bar rotation rate of $R_c/R_1 = 1.3^{+0.3}_{-0.2}$.

Figure~\ref{fig:dehnen_simu} shows the surface density (left column), the radial velocity (middle column) and the residual tangential velocity maps for the B5 and KRATOS simulations (top and bottom rows, respectively). The map of the residuals has been obtained by subtracting the rotation curve to the \vp\ map. In the surface density plot, the B5 simulation shows a strong bar accompanied by two strong spiral arms. In the KRATOS simulation, we observe a strong bar accompanied by a broken interacting arm. The radial and residual tangential velocity maps show the characteristic kinematic imprint of the bar, namely a quadrupole pattern. Larger velocities are observed  in the B5 simulation than in the KRATOS simulation (see also Sect.~\ref{subsec:simul_fourier}). 

We show first the results when applying the Dehnen method (Sect.~\ref{subsec:simul_dehnen}), because it defines the bar region in the simulations, which are used by the TW method (Sect.~\ref{subsec:simul_tw}). Finally, the results corresponding to the BV method are presented in Sect.~\ref{subsec:simul_fourier}.

\subsection{Results of the Dehnen method}
\label{subsec:simul_dehnen}

\citet{dehnen23} made their numerical tool to infer some bar properties from a single simulation snapshot available to the community. We thus used the Python code they provide, and, for both simulations  we have set various parameters. We fixed a minimum and maximum number of particles in the radial bins to $10^4$ and $5\times 10^4$ (respectively) for the B5 and KRATOS simulations. We adopted a maximum size of the sampling of the radial bins of 1.25 kpc, a minimum ratio of the strength of the surface density in the bar region of $\Sigma_2/\Sigma_0 = 0.1$ ($\Sigma_0$ and $\Sigma_2$ being the amplitudes of the  axisymmetric and bisymmetric surface density components), a maximum angular width of the bar of 10\degr, a minimum size of the bar of 1.25 kpc, respectively, with a minimum required number of particles in bar region of 1000. Following \citet{dehnen23} recommendations, we assumed a top-hat weighting function to constrain $\Sigma_2/\Sigma_0$ and $\phi_b$ in each bin from the surface density, and a smooth window to estimate \omp\ and the bar orientation in the entire bar region. We refer the reader to \citet{dehnen23} for more complete information of the code. The application of the Dehnen method to a simulated disc with a barred perturbation and no spiral structure is given in App.~\ref{sec:app-tpsim}.
 
\begin{figure*}
    \centering
    \includegraphics[width=1\textwidth]{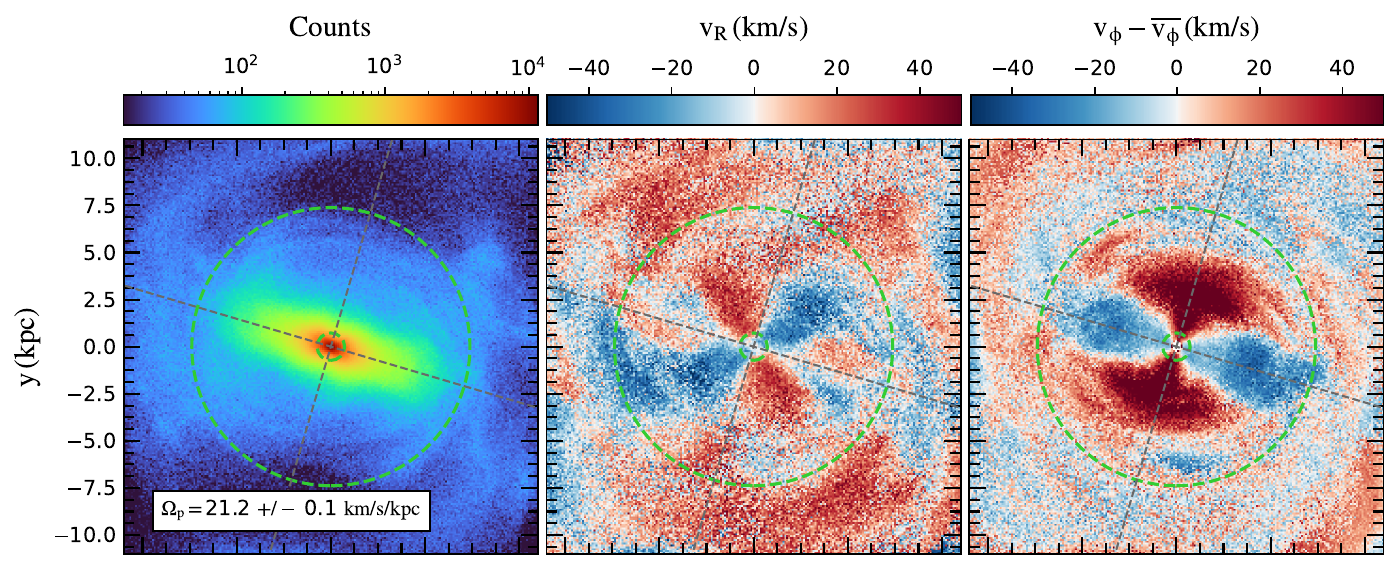}
    \includegraphics[width=1\textwidth]{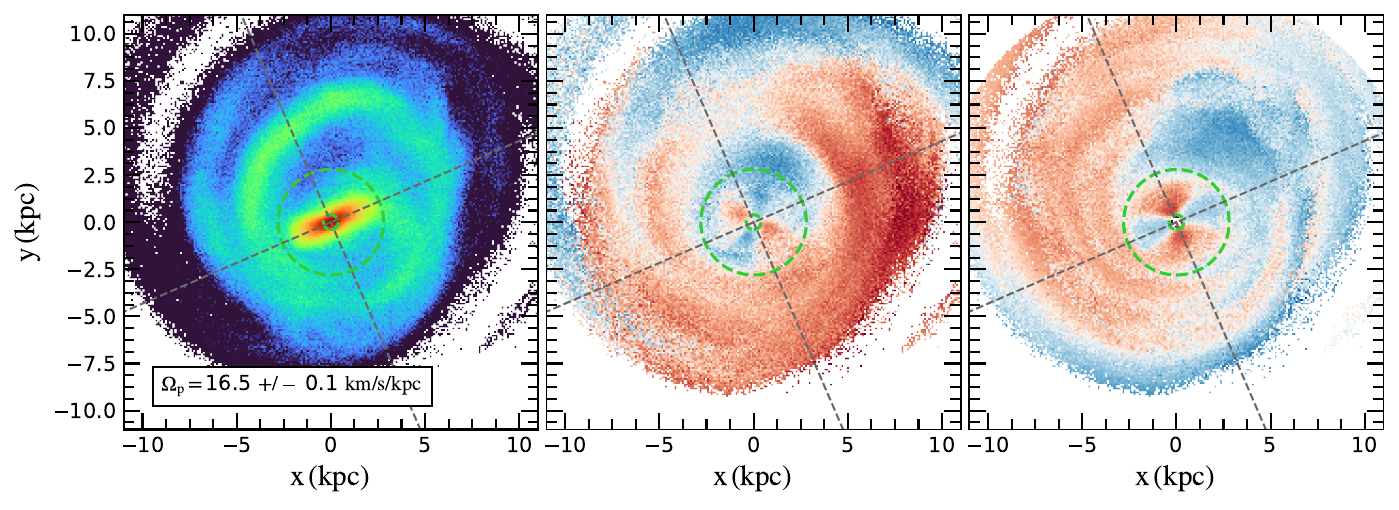}
    \caption{Application of the Dehnen method to B5 (top) and KRATOS (bottom) simulations. Surface density (left), median radial velocity map (centre) and median residual tangential velocity map (right). The bar region identified by Dehnen method (see values in Table~\ref{tabl:simulations}) is indicated by green dashed circles. The grey dashed lines trace the bar minor and the major axes.}
    \label{fig:dehnen_simu}
\end{figure*}

Figure~\ref{fig:dehnen_simu} shows the performance of the Dehnen method with both simulations. Left, centre and right panels show the surface density, the median radial velocity map and median residual tangential velocity map, respectively. In every panel, we highlight the bar region identified by the method by green dashed circles, with  inner and outer circles corresponding to $R_0$ and $R_1$, respectively. The grey dashed lines trace the bar minor and the major axes found by the method. For both simulations $\phi_b$ is in agreement with the orientation observed in the surface density, and separates remarkably  the quadrupole patterns in two parts. For the B5 simulation (top panels), the method infers a value of $\Omega_p = 21.2\pm 0.1$ \kmskpc, in good agreement with the value found using finite-differences. For the KRATOS simulation (bottom panels), the method infers a value of $\Omega_p = 16.5 \pm 0.1$ \kmskpc, also in  agreement with the value obtained using finite-differences. Values are summarised in Table~\ref{tabl:simulations}. These tests are thus another way to validate the method, in agreement to those performed in \citet{dehnen23}. It can be concluded that, under ideal conditions in which data are devoid of observational and numerical noise, the Dehnen method successfully recovers the imposed values of \omp.

This method yields a corotation radius of 8.3 kpc for the B5 simulation, corresponding to the ground-truth value. It gives a corotation radius of 4.0 kpc for KRATOS, which is very close to the ground-truth value.

\begingroup

\setlength{\tabcolsep}{10pt} % Default value: 6pt
\renewcommand{\arraystretch}{1.5} % Default value: 1
\begin{table*}
\centering
\begin{tabular}{llllllll}
\hline \hline
% \toprule
\multicolumn{1}{c}{} & \multicolumn{1}{c}{Reference} & \multicolumn{4}{c}{Dehnen method} & \multicolumn{2}{c}{BV method} \\
\cmidrule(rl){2-2} \cmidrule(rl){3-6} \cmidrule(rl){7-8}
     % & Reference &   &  &  &  \\
% \hline
 Simulation               & $\Omega_p$ &  $R_0$  & $R_1$ & $\Omega_p$ & $\phi_b$ & $R_c$ & $\Omega_p$ \\ \hline
B5    &  21.5 $\pm$ 0.1  & 0.73 & 7.38 & 21.2 $\pm$ 0.1 & 163.5 $\pm$ 0.1 & 8.0 $\pm$ 0.5 & 22.2$^{+0.7}_{-1.2}$ \\
KRATOS   &  17.2 $\pm$ 1.6  & 0.40 & 2.80 & 16.5 $\pm$ 0.1 & 23.4 $\pm$ 0.1 &  3.3 $\pm$ 0.5 &  18.2$^{+2.9}_{-1.3}$ \\
\hline \hline
% \bottomrule
\end{tabular}
\caption{Results of the Dehnen and the BV methods applied to B5 and KRATOS simulations, compared to the reference value (obtained using finite-differences). The inner, outer and corotation radii $R_0$, $R_1$ and $R_c$ are in kpc. The bar pattern speed \omp\ and phase angle $\phi_b$ are in \kmskpc and degrees, respectively.}
\label{tabl:simulations}
\end{table*}

\endgroup

\subsection{Results of the TW method}
\label{subsec:simul_tw}
As mentioned above, the applicability of the TW method in either of the two versions is designed for galaxies in equilibrium and, moreover, with a single pattern speed. We can nevertheless evaluate their performance by applying them to the simulations of isolated and interacting spiral discs. Application of the methods to a simulated disc with a barred perturbation and no spiral structure is given in App.~\ref{sec:app-tpsim}.

Hereafter, the domain [$-\infty/+\infty$]   of the numerical integration of Eqs.~\ref{eq:tw_vxvy} and~\ref{eq:tw_vlos} is the maximum extent of the disc allowed inside each pseudo-slit parallel to the $x-$axis at a given $y$ coordinate, or the line-of-nodes for the LTW method. Within this maximum range, the integrals have converged to stable values. Furthermore, we investigated the impact of the width of  the  pseudo-slits. Within a range of 50-500 pc, we found that the width has no effect on the results described below. Only the uncertainties on \omp\ are seen to increase for wider pseudo-slits, by a factor of 3.5 from 50 pc to 500 pc width. In this Section, we show the results obtained for a width of 200 pc.

\subsubsection{Results of the LTW method}
\label{subsec:simul_tw_vlos}

We simulate a galaxy observation by projecting the particles  from each simulation onto a galactic plane with arbitrary inclination  ($i$) and position angle (PA) of the semi-major axis of the receding half. For the result described here, we   adopted a disc projected with $\rm PA = 60\degr$, and inclinations of $i=25\degr$, $45\degr$ and $75\degr$.  Assuming a mock disc distance of 10 Mpc, which is well suited to mock galaxies on which the LTW method can been applied, the maps of the projected density and \los\ velocity have $512 \times 512$ pixels sampled at 1\arcsec\ ($\sim 50$ pc/pixel). We do not investigate the impact of varying  the distance of the mock disc on the results. We chose the $x-$axis aligned with the line of nodes, so that the reference of the azimuthal angle $\phi = 0$ is along the semi-major axis of receding disc half. The small angle approximation can be applied, and the \los\ kinematics of each particle simply resumes to $V_{\rm los} = V_z \cos i + V_y \sin i = V_z \cos i + (V_R \sin \phi  + V_\phi \cos \phi) \sin i$. The adopted velocity in each pixel of the map is the mean of the \vlos\ of the particles.  The LTW integrals of Eq.~\ref{eq:tw_vlos} are  performed at each $Y-$coordinate, selecting all pixels from the maps within pseudo-slits parallel to the disc major axis. The derivation of the slope $\Omega_p \sin i$ is performed   using the  $\langle X \rangle$-$\langle V_{\rm los} \rangle$ points located inside the region encompassing the projected value of the bar radius $R_1$ from in Table~\ref{tabl:simulations}.

\begin{figure}
    \centering
    \includegraphics[width=\columnwidth, clip,trim=-0.1cm 0.3cm 0cm 0cm]{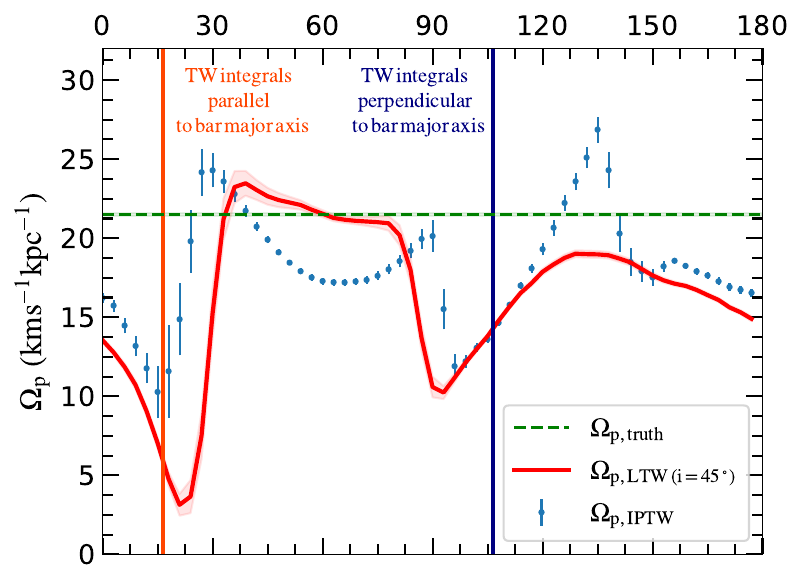}
    \includegraphics[width=\columnwidth, clip,trim=0.3cm 0cm 0cm 0.25cm]{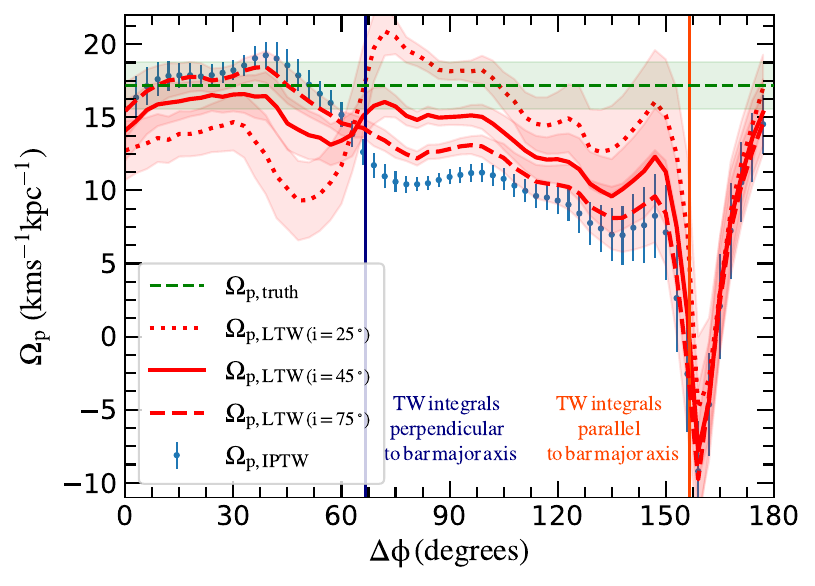}
    \caption{Variation of \omp\  as as function of the reference frame orientation $\Delta \phi$. Results for the B5 and KRATOS simulations are shown in the upper and lower panels, respectively. Results of the IPTW method are shown as open symbols, while those of the LTW method are drawn as a solid line (for the $i=45\degr$ case, upper panel), and as dotted, solid and dashed lines (for the $i=25, 45$ and $75\degr$ cases, lower panel). Horizontal dashed green lines are the ground-truth bar pattern speeds. The vertical orange (navy) vertical line corresponds to the frame orientation where the $x$-axis of the disc lies along the major axis of the bar ($y$-axis, respectively).}
    \label{fig:ltw_iptw}
\end{figure}

We  investigate the impact of the variation of the orientation of the Cartesian frame on the results, by rotating the reference $x$ and $y$ axes  around the $z-$axis in the simulation. The rotation of the Cartesian frame before projection on the sky plane allows the TW integrals to view the bar and spiral perturbations through various angles. In practice, this is achieved by adding a $\Delta\phi$ to the  angular position $\phi$ of each particle in the disc plane, from which new $x$ and $y$ positions are derived. Then, new density maps and \los\ kinematics and \omp\ can be inferred.  A range of $\Delta\phi$  spanning $180\degr$ has been probed, with a step of 3\degr. Figure~\ref{fig:ltw_iptw} (upper panel) shows the resulting bar pattern speed as a function of the frame orientation $\Delta\phi$ for the case $i=45\degr$ only (red open symbols) and the B5 simulation. The quoted uncertainties correspond to the 1$\sigma$ error of the covariance matrix of the fitting. We highlight  the frame orientations parallel and perpendicular to the bar major axis as orange and navy vertical lines, respectively. Results for the KRATOS simulation are shown in the lower panel of  Fig.~\ref{fig:ltw_iptw} for the three assumed inclinations.
 
The variation of the pattern speed with $\Delta\phi$  is very important in both simulations, and makes it  rarely consistent with the ground-truth values (shown as horizontal green dashed lines), whatever the adopted disc inclination. The strongest disagreement occurs at frame orientations very close to the major axis of the bar, and $\sim 15\degr$ before its minor axis. For the B5 simulation, the LTW results  are consistent with the real value only at frame orientations $\Delta\phi \sim 70-75\degr$, thus when the  $x-$axis in the galaxy plane makes an angle of $\sim 55-60\degr$ with respect  to the bar major axis.  The  LTW pattern speeds agree with the ground-truth value only occasionally for the KRATOS simulation, within the quoted uncertainties (shown as shaded areas). A strong dependency with the inclination of the disc is observed. The best agreement with ground-truth  is for a mock disc at $i=45\degr$, for $7\degr < \Delta\phi < 38\degr$, thus when the  $x-$axis in the galaxy plane makes an angle of $\sim 30-60\degr$ with respect to the bar major axis. Another interesting result is that no symmetry around the bar minor axis is observed, indicating that having a wide range of agreement with ground-truth is unlikely (see Sect.~\ref{subsec:simul_tw_vxvy} for more details).
 
\subsubsection{Results of the IPTW method}
\label{subsec:simul_tw_vxvy}

\begin{figure*}
    \centering
    \includegraphics[width=0.8\textwidth]{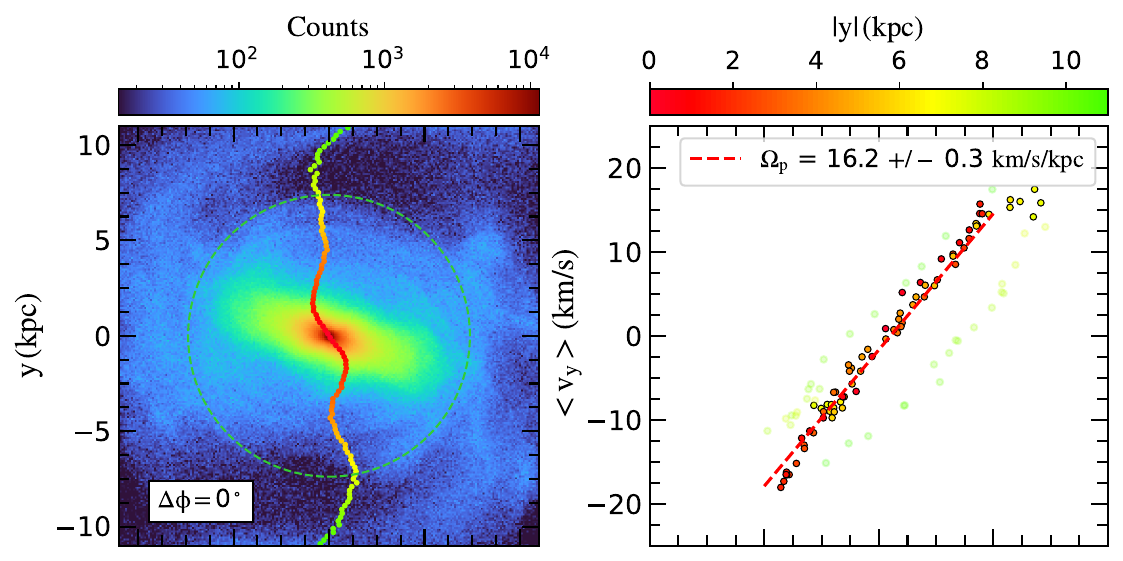}
    \includegraphics[width=0.8\textwidth]{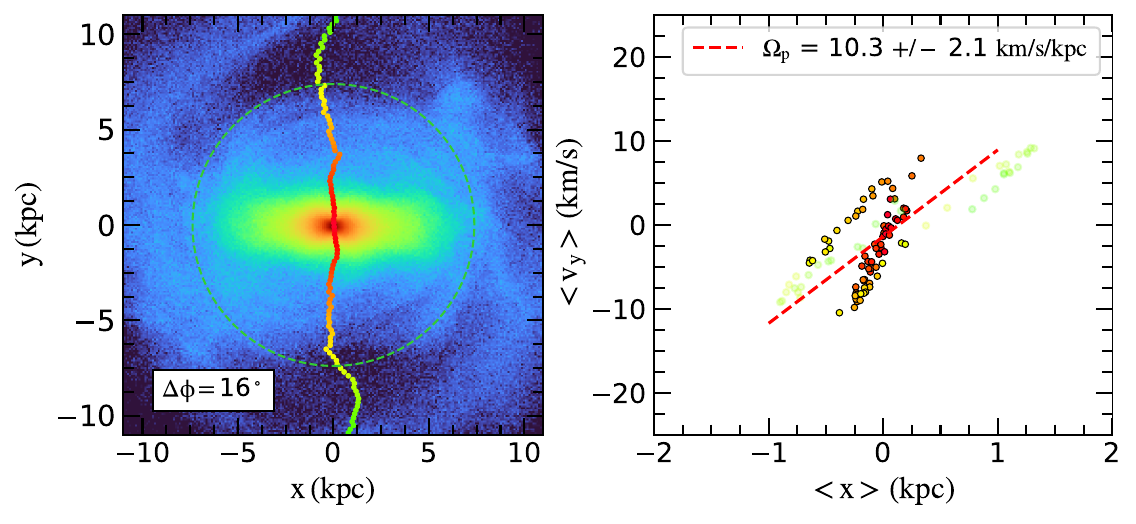}
    \caption{Application of the IPTW method to B5 simulation. Top: The bar major axis at $\phi_b=163.5$\degr, as in the original snapshot and shown in Fig.~\ref{fig:dehnen_simu}. Bottom: After applying a rotation of $\Delta \phi=16.5$\degr to put the bar major axis along the $x$-axis. Left: Surface density where the bar region (obtained using Dehnen method) is indicated by the green dashed circle. A scatter plot representing the value of $\langle x \rangle$ for each slice in $y$ is overlapped. The scatter plot varies its colour as function of the distance to the centre in the $y$-axis, being the red points close to the centre and the green close to the external parts of the galaxy. Right: Scatter plot of the Tremaine-Weinberg integrals $\langle x \rangle$ and $\langle v_y \rangle$ for the different slices in the $y$-axis. The colour of the scatter plot is the same in both left and right panels. In the right panel, the points with a black circle are inside of the bar region and, therefore, the only considered for fitting $\Omega_p$.}
    \label{fig:tw_B5}
\end{figure*}

As in the LTW method, we varied  $\Delta\phi$ between 0\degr\ and 180\degr\ to study the dependency of the in-plane TW  integrals of Eq.~\ref{eq:tw_vxvy} with the frame orientation. Figure~\ref{fig:tw_B5} shows  results for the B5 simulation for two examples of  frame orientations\footnote{Animations of the variation of the pattern speed $\Omega_p$ inferred by the IPTW with different frame orientations $\Delta\phi$ are available online, for both simulations and data.}. The upper row corresponds to the original frame orientation, with $\Delta\phi=0$\degr, while in the bottom row, the $x$-axis is chosen aligned with the bar major axis,  corresponding to a frame orientation of 16.5\degr, which  corresponds to $\Delta\phi=16$\degr. The left panels show the surface density maps, with the bar region outer radius $R_1$ identified with the Dehnen method highlighted by a green dashed circle. The coloured dots represent the $\langle x \rangle$ integrals for each slice in $y$, with greener (redder) colours for larger (smaller) values of $|y|$. In the case where the bar major axis is parallel to the $x$-axis, $\langle x \rangle$ is observed close to the $x=0$ axis in the bar region, thus rather aligned with the bar minor axis. In the case where the bar major axis is 30\degr\ rotated counter-clockwise with respect to the $x$-axis, $\langle x \rangle$ varies significantly in the bar region, almost tracing the bar major axis. 

The right panels show the TW integrals, $\langle x \rangle$ versus $\langle v_y \rangle$, with the same colour code for the various $y$ as those in the left panels.  The points with a black circle are those located inside $R_1$, while the rest of the points are the ones outside the bar region, therefore, not considered for fitting  \omp. In the case with $\Delta\phi=0$, thus the original frame orientation, there is a clear linear trend with small dispersion for points inside the bar region, as shown by the dashed straight line fit. In the case where the bar major axis is parallel to the $x$-axis, several linear trends are observed, hence a significantly larger dispersion. The fitted slope of the TW integrals is $10.3 \pm 2.1$ \kmskpc, as shown by a dashed straight line. However, none of these values  are consistent with the real pattern speed of $21.5 \pm 0.1$ \kmskpc, or the value of $21.2 \pm 0.1$ \kmskpc\ from the Dehnen method.

\begin{figure*}
    \centering
    \includegraphics[width=0.8\textwidth]{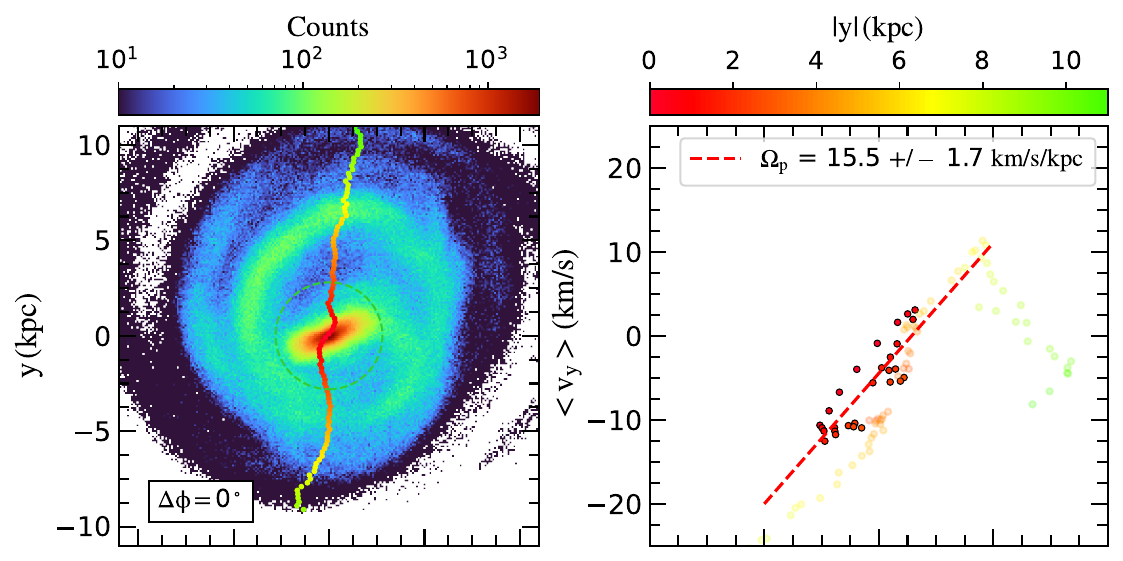}
    \includegraphics[width=0.8\textwidth]{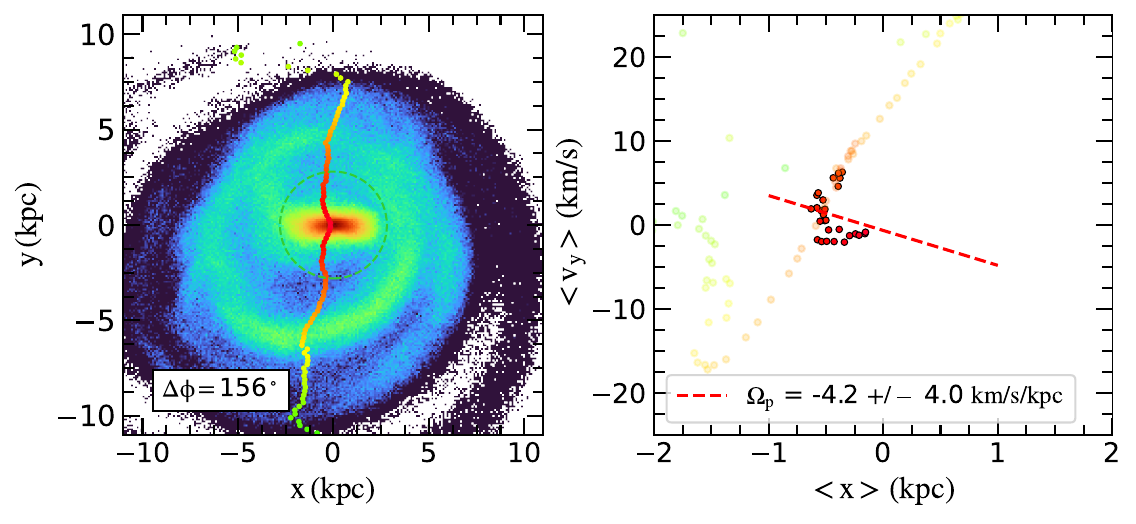}
    \caption{As in Fig.~\ref{fig:tw_B5} for the KRATOS simulation. }
    \label{fig:tw_KRATOS}
\end{figure*}

Figure~\ref{fig:tw_KRATOS} presents results of the IPTW method for the KRATOS simulation. Again, we show here two different reference frame orientations: the original one, with $\Delta\phi=0$\degr\ (top row) and the one with the bar major axis parallel to the $x$-axis, $\Delta\phi=156$\degr\ (bottom row). In the first case, there is a linear trend and the IPTW recovers a bar pattern speed $\Omega_p = 15.5 \pm 1.7$ \kmskpc, compatible with the real pattern speed. In the second case, the TW $x-$integrals are perfectly aligned with the bar major axis, as expected in the presence of only a bar potential. The $\langle x \rangle$ values are then very close to zero, so there is not a clear trend in this case (it would give similar results when the TW integrals are evaluated at another viewing angle along the bar minor axis). We recover here a clockwise pattern speed of $\Omega_p = -4.2 \pm 4.0$ \kmskpc\ which differs from the real value. Note also that in this configuration, the linear trend is more scattered, not as clear as the configuration from the upper panel, in agreement with what was seen with for the B5 simulation. 

We then assess whether there is (or not) a favoured frame orientation where the IPTW method works better, by plotting the fitted \omp\ as a function of $\Delta\phi$ (filled blue symbols in Fig.~\ref{fig:ltw_iptw}). Qualitatively, the IPTW and LTW methods show similar trends:  the agreement with the ground-truth bar pattern speed is rarely observed, no symmetry with respect to the bar major axis is found, and  stronger discrepancies are near the positions of the major and minor axes of the bar. The B5 simulation (upper panel) shows that $\Delta\phi \sim 135\degr$ is also a location of stronger disagreement, which was not observed for the LTW method. Since the IPTW method works directly with coordinates and velocities in the Cartesian frame of the disc, no variation with inclination needs to be evaluated here. It is interesting to note that the LTW pattern speeds with better agreement with the IPTW method are for the intermediate inclination of 45\degr\ (lower panel for the KRATOS simulation). It is important to remind that the Dehnen method does not show such systematic variation with $\Delta\phi$, as its results are invariant with the frame orientation.

The median and mean absolute deviation of all the IPTW values are $17.9\pm 1.9\,$\kmskpc\  and $11.0\pm 3.7\,$\kmskpc\ for the B5 and KRATOS simulations, respectively. We note that the difference is larger in the KRATOS simulation. While the B5 simulation presents a strong spiral pattern in the outer disc, the KRATOS simulation additionally is not in equilibrium.

Finally, we can estimate the incidence of finding a bar pattern speed consistent with the ground-truth value for both the LTW and IPTW methods, with the two simulations. We define this likelihood as the number of frame orientations where the measured  and real \omp\ agree within the quoted ($1\sigma$) uncertainties on measured and ground truth values. For the B5 simulation, the  IPTW and LTW methods give a correct \omp\ in 5\% and 8\% of the cases only. For the KRATOS simulation, the incidence is 37\%  (IPTW case), 57\% (LTW case at $i=25\degr$), 48\% (LTW case at $i=45\degr$) and 42\% (LTW case at $i=75\degr$). Larger inclinations are thus less prone to the LTW method. More generally, our two sets of simulations show it is highly unlikely to find a consistent \omp\ by means of the TW method. It is also hard to reconcile the strong variations with  $\Delta\phi$ seen here, i.e the bar orientation with respect to the disc $x-$axis, with the wide range of ``allowed'' orientations quoted in other studies \citep[e.g.][see also Sect.~\ref{sec:discussion} for the LMC]{2019zou}.

\subsection{Results of the bisymmetric velocity model}
\label{subsec:simul_fourier}

Bayesian inferences of Fourier coefficients to the tangential velocities were performed in radial bins through Markov Chain Monte Carlo   fits, using the Python library Emcee \citep{2013foreman}. The model is fitted to a map of $V_\phi$ (pixel size of 50 pc), where the velocity at each pixel of the map is the median of  the velocity distribution from all particles/stars inside the given pixel.  Defining the residual velocity as $V_{\phi,\text{res}} = V_\phi - V_{\phi,\text{mod}}$, the conditional likelihood function at each radial bin is expressed by:
\begin{equation}
     \mathcal{L} (V_{0}, V_2, \phi_{2,\rm kin}, V_{s}) = - \frac{1}{2} \left( n_{\text{pix}} \ln (2 \pi) \sum^{n_{\text{pix}}} \left( V_{\phi,\text{res}}^2 / \xi^2 + \ln (\xi^2)  \right)  \right),
\end{equation}
where $V_{0}, V_2$ and $\phi_2$ are as in Eq.~\ref{eq:modbisym}, $\xi^2 = \sigma^2_{V_\phi} + V^2_{s}$, $\sigma_{V_\phi}$ are the $V_\phi$ uncertainties, $V_{s}$ is the scatter of the modelling, and $n_{\text{pix}}$ the number of pixels of the map inside the corresponding radial bin at which parameters are fitted. No uncertainties are measured while making the velocity map, thus $\sigma_{V_\phi}=0$. Therefore, measuring $V_{s}$ is an indirect way to take into account the lack of uncertainties \citep{2010hogg}. 

\begin{figure*}
    \centering
    \includegraphics[width=0.48\textwidth]{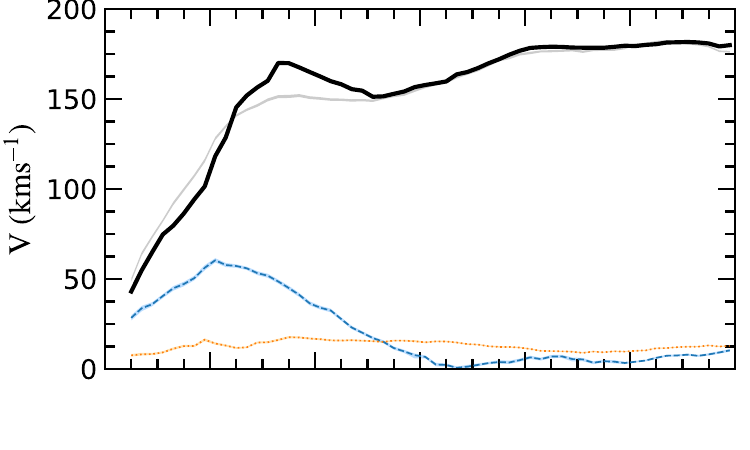}
    \hspace{-0.2cm}
    \includegraphics[width=0.48\textwidth]{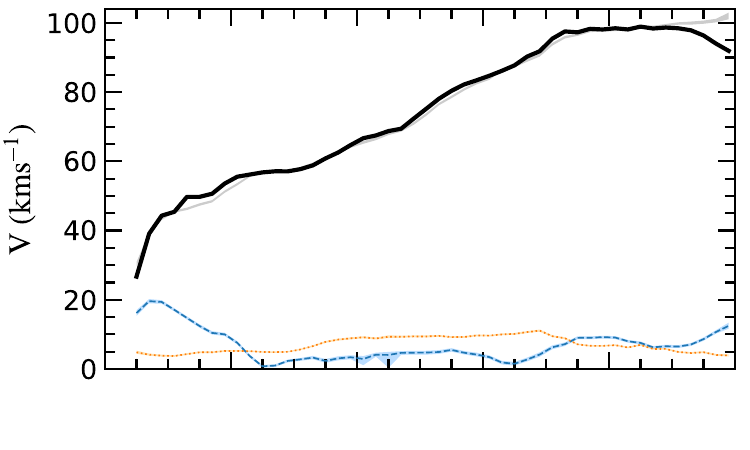}\\
    \vspace{-0.75cm}
    \includegraphics[width=0.48\textwidth]{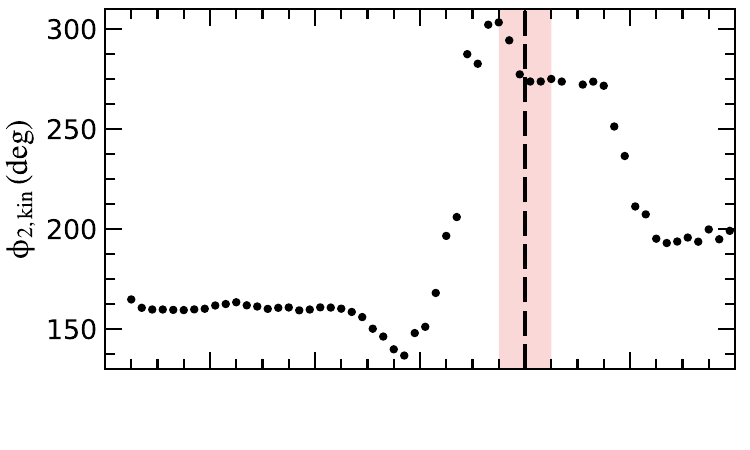}
    \hspace{-0.2cm}
    \includegraphics[width=0.48\textwidth]{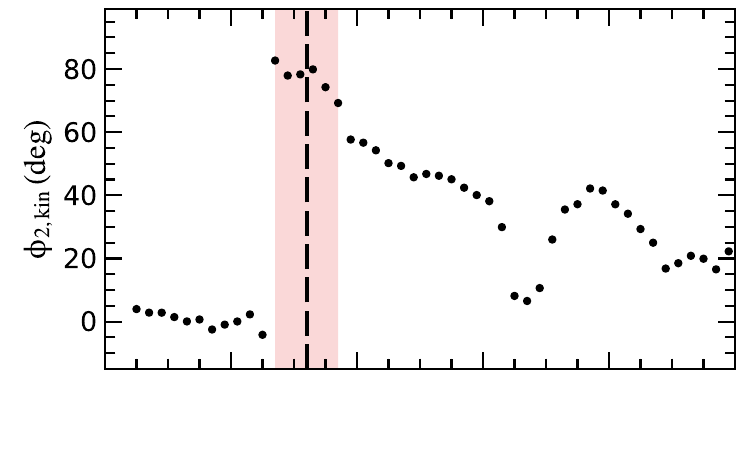}\\
    \vspace{-0.75cm}
    \includegraphics[width=0.48\textwidth]{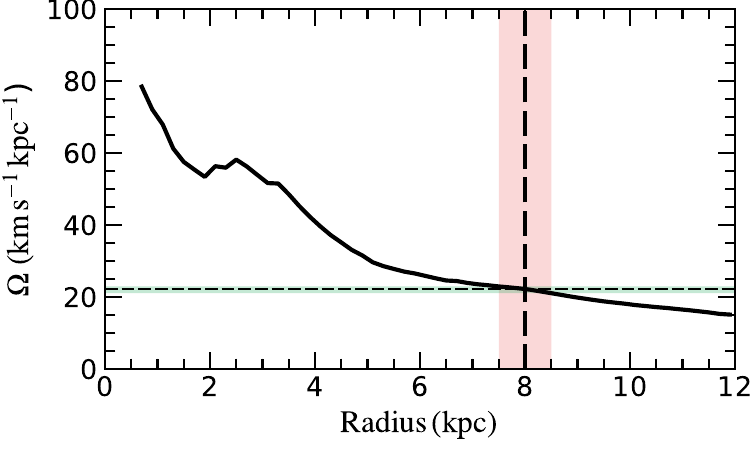}
    \hspace{-0.2cm}
    \includegraphics[width=0.48\textwidth]{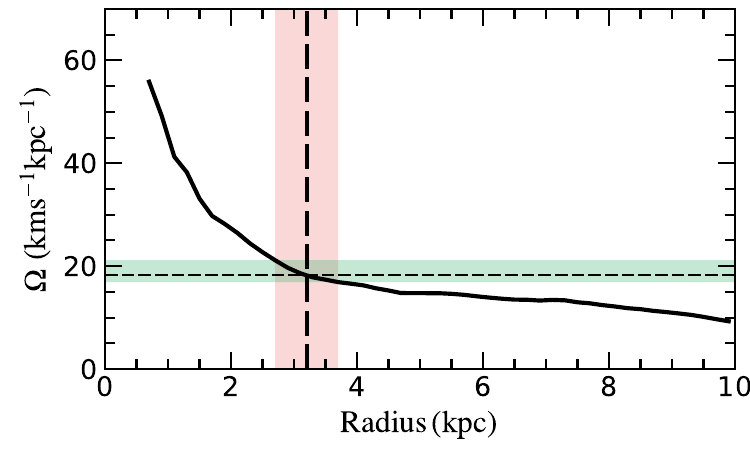}
    \caption{Results of the BV model of the stellar tangential velocity map of the B5 (left panels) and KRATOS (right panels) simulations: amplitude (upper panels), phase angle, $\phi_{2,\rm kin}$ of the Fourier modes (middle panels), and angular frequency $\Omega$ (bottom panels). In the top panels, the black solid line is the fitted axisymmetric velocity component $V_0$ (the rotation curve), the grey line is the median velocity (initial value for the model), the blue dashed curve is the amplitude of the tangential bisymmetry $V_2$, and the orange dotted line is the scatter  of the model $V_{s}$. In the middle and bottom panels, the vertical light coral area and dashed line shows the adopted bar corotation radius, $R_c = 8.0 \pm 0.5$ kpc for B5 and $R_c = 3.3 \pm 0.5$ kpc for KRATOS simulations. In the bottom panel, the black solid line is the angular velocity derived from $V_0$, while the horizontal green area and dotted line shows the corresponding bar pattern speed $\Omega_p = 22.2_{-1.2}^{+0.7}$ \kmskpc\ and $\Omega_p = 18.2_{-1.3}^{+2.9}$ \kmskpc, for the B5 and KRATOS simulations, respectively. Values are summarised, and compared with the reference values, in Table~\ref{tabl:simulations}.}
    \label{fig:fourier_simu_corotation}
\end{figure*}

The left column of Fig.~\ref{fig:fourier_simu_corotation} shows the results of the model to the B5 simulation. In the top panel, we show the   rotation curve of the simulated disc (gray solid line, measured as the median velocity in the map a each $R$), the fitted axisymmetric velocity component (black solid line, $V_{0}$),  the amplitude of the bisymmetry $V_2$ (blue dashed line), and the scatter in  $V_{\phi,\text{res}}$  (orange dotted line, $V_{s}$). The bar strength is maximum at $R\sim 2$ kpc, reaching more than 50\% of $V_{0}$.  Within $R=5$ kpc,  $V_{0}$ and the median rotation curve can differ by up to $\sim 20$ \kms, which indicates the significant impact of the bisymmetry on the rotation curve. In the middle panel, we show the phase angle $\phi_{2,\text{kin}}$ of the bar recovered from the modelling of the tangential velocity map. The phase angle of the bar is well recovered by the bisymmetric model at low radius, then smoothly varies. At $R\sim 7.5$ kpc, an abrupt change in $\phi_{2,\text{kin}}$ is observed. Following prescriptions given in Sect.~\ref{sec:methods}, we can identify the  corotation radius just after the steep change of phase angle, $R_c = 8.0 \pm 0.5\,$kpc, which corresponds to a bar pattern speed is $\Omega_p = 22.2_{-1.2}^{+0.7}$ \kmskpc\ (bottom panel, the $\Omega$ curve being derived from the solid curve of the upper panel, i.e. the 0th order Fourier coefficient). This agrees  with the value computed using finite-differences. By construction, the  bisymmetric velocity model is invariant with the frame orientation, as $\phi_{2,\text{kin}}$ are shifted by $\Delta\phi$ when a rotation of $\Delta\phi$ is applied to the $x-y$ plane.

Similarly, in the right panels of Fig.~\ref{fig:fourier_simu_corotation}, we show the results of applying this method to the KRATOS simulation, the value for corotation we derive is $R_c = 3.3 \pm 0.5\,$kpc, and the bar pattern speed is $\Omega_p = 18.2_{-1.3}^{+2.9}$ \kmskpc, which is consistent with the true value of $17.2 \pm 1.6$ \kmskpc, within the quoted uncertainties. Values are summarised in Table~\ref{tabl:simulations}.

This method yields a bar rotation rate of $R_c/R_1 = 1.1$ and $1.2$ for the B5 and KRATOS simulations, respectively, in agreement with the ground-truth values.

\section{Measuring the LMC bar pattern speed}
\label{sec:results}

In the previous section, we have assessed the performance, robustness and limitations of the methods when applied to simulated data. The TW method, in both versions using either line-of-sight or in-plane velocities, shows a clear dependence of the measured pattern speed on the frame orientation, and significant differences with the ground-truth values. The recovered value can be lower or higher than the simulated speed when both bar and spiral arms intervene in the simulation.  These results already warn us to take the value of the LMC pattern speed inferred from the TW method with caution.

The  LMC stars we use in this study are those in the NN complete sample of \citet{jimenez-arranz23}.  The selection of LMC stars was based on a supervised neural network classifier, using full astrometric and photometric data from \gaia DR3. Based on this classifier, the authors select three samples of candidate LMC stars with different degrees of completeness and purity. The NN complete sample corresponds to the sample that prioritises not missing LMC stars at the price of a possible increased MW contamination. It contains 12~116~762 stars. The sample is dominated by older stellar populations \citep[see, e.g., Fig.3 from][]{luri20}, thus fulfils the continuity equation necessary to the TW and Dehnen methods. Combining the selection function of the \gaia parent catalogue \citep{2023A&A...669A..55C} and the selection effects from the generation of the LMC NN complete sample \citep[estimated as in][]{2023arXiv230317738C}, the completeness estimates of our sample are above $50\%$ in the bar region for $G = 19 - 19.5$ mag. As seen in Fig.~6 of \citet{jimenez-arranz23},  60\% of stars in the LMC NN complete sample have magnitude below $G<19.5$. To better estimate the completeness and purity of the LMC NN complete sample, and their effect on the inner kinematics, a more detailed study of the selection function is required, which is out of the scope of this paper. 

For each star, we apply the coordinate transformation detailed in \citet{jimenez-arranz23} to express the deprojected positions and velocities in the in-plane coordinate system of the LMC, using the inclination,  position angle,  systemic velocity, and position of the LMC centre given in \citet{jimenez-arranz23}, assuming all stars lying in the $z=0$ plane. The LMC centre used is the same as in \citet{gaiadr2,luri20}, which corresponds to the LMC photometric center \citep{vandermarel01}. The infinitely thin disc approximation is inherent to any studies of the kinematics of disc galaxies because the 3D position space of stars in galaxies is never available, unlike stars in the MW \citep[e.g.][]{drimmel22}, or a few variable young  stars in the LMC \citep{2022ripepi}. 

\begin{figure*}
    \centering
    \includegraphics[width=1\textwidth]{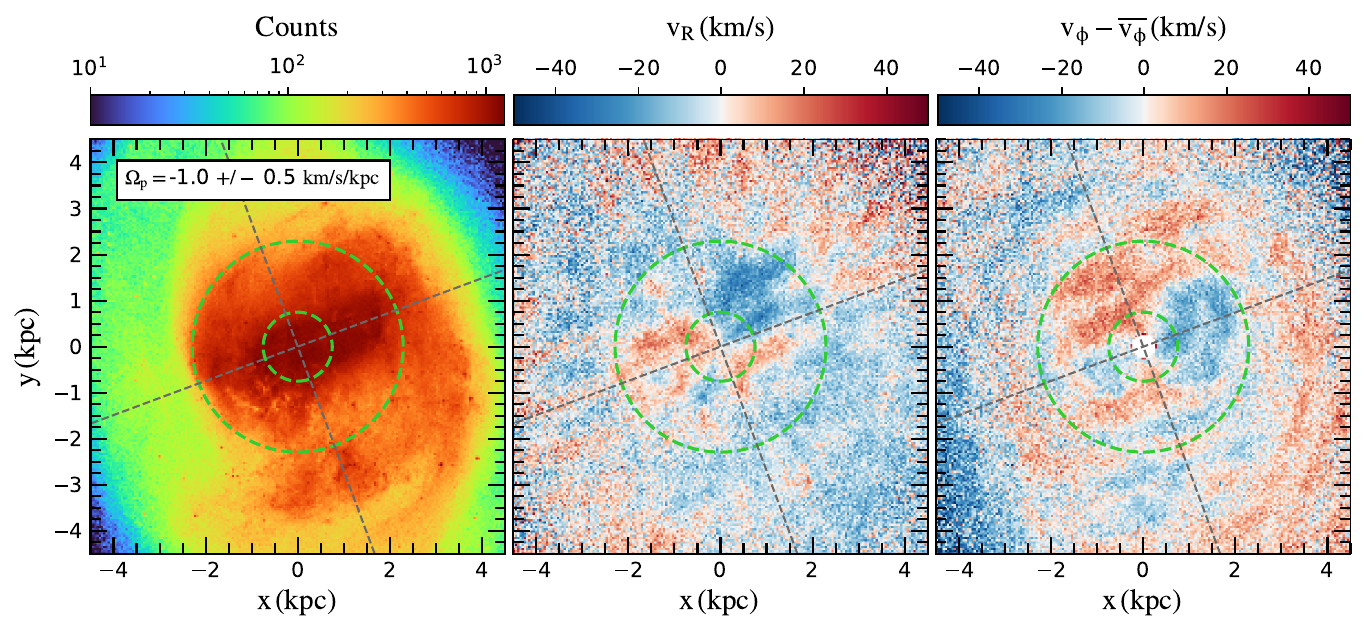}
    \caption{Application of the Dehnen method to the LMC NN complete sample. Surface density (left), median radial velocity map (center) and median residual tangential velocity map (right). The bar region identified by Dehnen method is indicated by green dashed circles, $[R_0, R_1]=[0.75,2.3]\,$kpc. The grey dashed lines trace the bar minor and the major axes.}
    \label{fig:dehnen_lmc}
\end{figure*}

\begin{figure}
    \centering
    \includegraphics[width=0.5\textwidth]{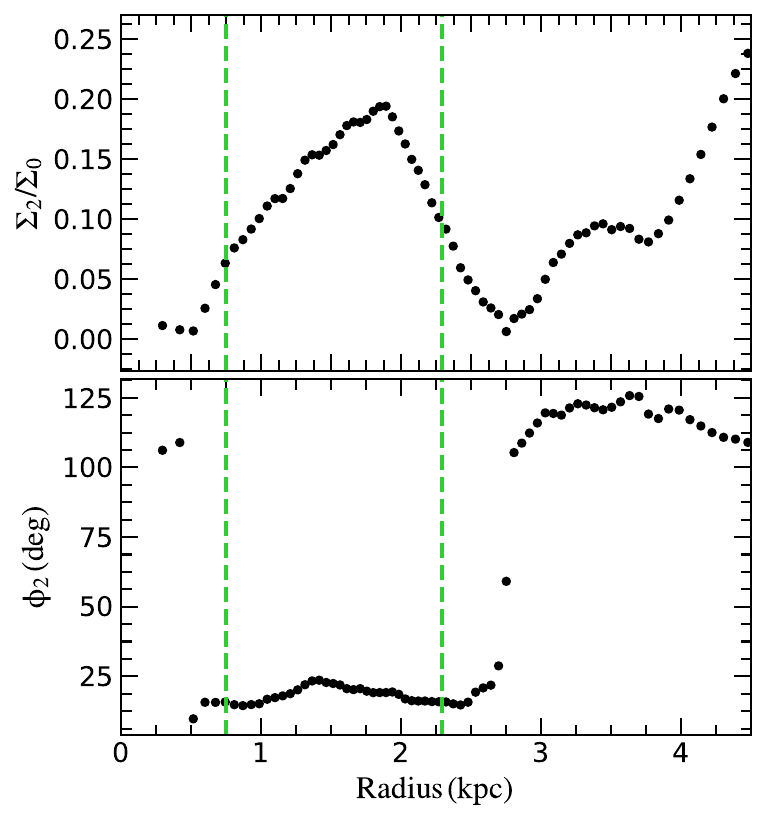}
    \caption{Application of the Dehnen method to the LMC NN complete sample. The relative $m = 2$ Fourier amplitude (top) and orientation (bottom). The bar region identified by the method is indicated by green dashed vertical lines, $[R_0, R_1]=[0.75,2.3]\,$kpc.
    }
    \label{fig:dehnen_lmc_amplitude}
\end{figure}

Figure~\ref{fig:dehnen_lmc} shows the results of the application of the Dehnen method to the LMC. The left, middle and right panels show the surface density, the median Galactocentric radial velocity, and  residual of the median tangential velocity, respectively. Both the radial and residual tangential velocity maps show the imprint of a rigidly rotating bar in the galactic centre, as clear hints of $x_1$ stellar orbits \cite[see also][]{luri20, niederhofer22,jimenez-arranz23}. The quadrupole pattern is the natural reflex of the motion of stars in elliptical orbits present in the bar potential. In contrast to the simulations, the method was not able to find the bar region $[R_0, R_1]$ on its own, probably due to the fact that the quadrupole is not perfectly symmetric, or that the contrast of the bar region with respect to the disc is not as clear as in simulations because of the presence of dust lanes and spiral arms at low radius. We solved this issue by analyzing the outputs of the numerical code of \cite{dehnen23} which, in addition to the bar parameters and pattern speed, provide results of a second order Fourier model of the stellar density. Figure~\ref{fig:dehnen_lmc_amplitude} shows the amplitude $\Sigma_2$ of the $m = 2$ Fourier coefficient, relatively to the axisymmetric density $\Sigma_0$, and the phase angle $\phi_2$ of the bisymmetric density perturbation. In the upper panel,  a peak of the relative strength $\sim 0.2$ is observed at $R \sim 2$ kpc. This amplitude is comparable to that at larger radius, meaning a small contrast between the LMC bar and the spiral arm(s) within the selected sample  of stars. This probably explains why the Dehnen method cannot establish properly the bar region  in its automated way. The phase angle (bottom panel of Fig.~\ref{fig:dehnen_lmc_amplitude}) has a constant value of $\phi_2 \simeq 15-20$º from $R = 0.75$ kpc to $R \sim 2.3$ kpc.  We can establish that the bar region is  theferore $[R_0, R_1]=[0.75,2.3]\,$kpc. 
Within this region, the Dehnen method gives a value of $\Omega_p = -1.0 \pm 0.5$ \kmskpc, thus corresponding to an almost non-rotating stellar bar, seemingly in counter-rotation. 

\begin{figure}
    \centering
    \includegraphics[width=\columnwidth]{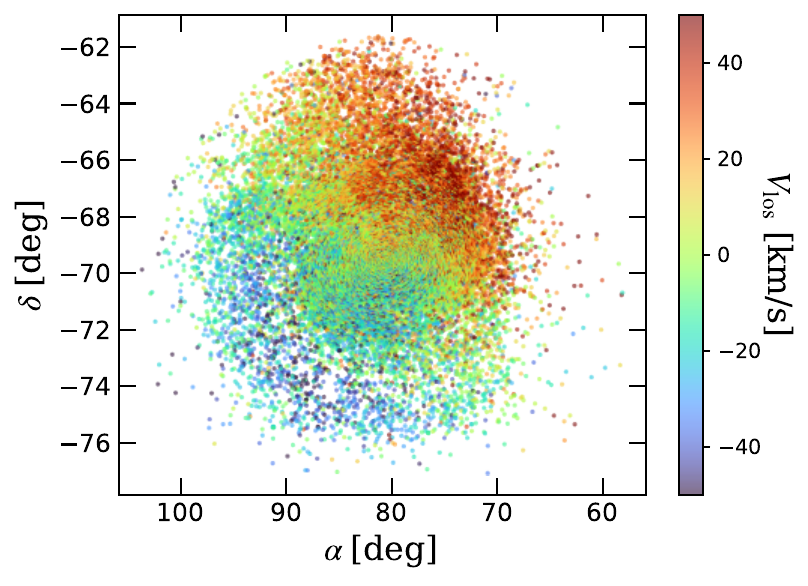}
    \caption{Stellar line-of-sight velocity field of the LMC NN complete $V_{los}$ sub-sample, corrected from the systemic motions. Data  are from \gaia RVS \citep{Katz2022,jimenez-arranz23}.}
    \label{fig:vflos_lmc}
\end{figure}

Now for the TW method, we adopted an pseudo-slit length and width of $[-\infty,+\infty]$ and 50 pc, respectively. For both versions of the method, the slopes of the integrals are fitted using only points located inside the bar radius (for the IPTW case) and projected radius (for the LTW case), as defined by $R_1  = 2.3$ kpc.

\begin{figure}
    \centering
    \includegraphics[width=\columnwidth]{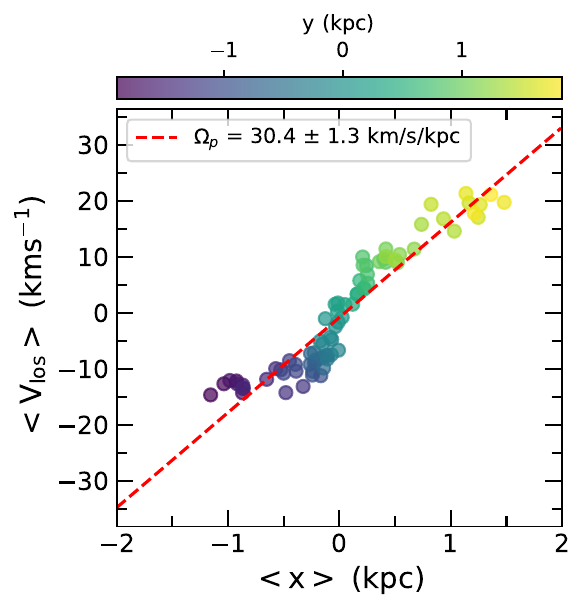}
    \caption{Result of the TW method applied to the \los\ velocity field of the LMC from Fig.~\ref{fig:vflos_lmc}. Only points within the radius $R_1$ defined by the Dehnen method are shown. The red dashed line shows the result of the linear fit to the points, $\Omega_p \sin i$, with $\Omega_p = 30.4 \pm 1.3$ \kmskpc.}
    \label{fig:ltw_lmc}
\end{figure}

In Fig.~\ref{fig:vflos_lmc}, we show the stellar \los\ velocity field corrected for the systemic motion of the LMC NN complete \vlos\ sub-sample, which contains 30~749 stars. These are predominantly the AGB stars from \citet{luri20}. This is in good agreement with the \los\ velocity field traced by carbon stars \citep{vandermarel02}. We apply the LTW method to this \los\ velocity map, as described in Sect.~\ref{subsec:method_tw}, with pseudo-slits parallel to the line-of-nodes. Figure~\ref{fig:ltw_lmc} shows the linear fit to the LTW integrals,  yielding a bar pattern speed of $\Omega_p = 30.4 \pm 1.3$ \kmskpc, using an inclination of $i=34\degr$.  

\begin{figure}
    \centering
    \includegraphics[width=0.5\textwidth]{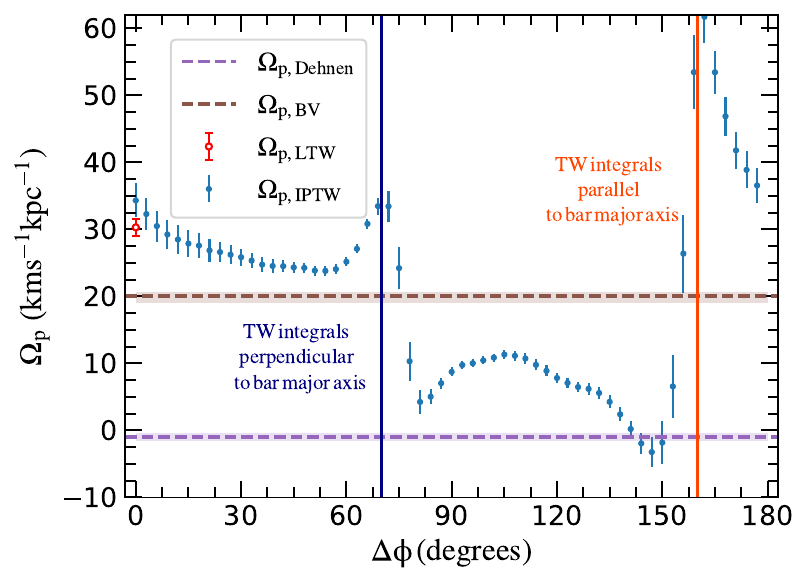}
    \caption{Results of applying the LTW and IPTW to the LMC NN complete sample. The open red dot shows the bar pattern speed recovered with the LTW method. The blue dotted curve shows the bar pattern speed obtained using the IPTW with different frame orientations $\Delta\phi$. The dashed purple (brown) line shows the bar pattern speed obtained using the Dehnen (BV) method. }
    \label{fig:tw_rot_LMC}
\end{figure}

Figure~\ref{fig:tw_rot_LMC} shows the results of the IPTW method applied to the LMC Cartesian velocity fields (not shown here, but obtained from the cylindrical velocities shown in Fig.~\ref{fig:dehnen_lmc}). We recover as many values as  adopted orientations $\Delta\phi$ of the Cartesian frame in the LMC plane. Interestingly, a good agreement is seen between the LTW \omp\ (open red dot at $\Delta\phi=0$\degr) and the IPTW \omp\ inferred at this orientation. However, and unsurprisingly, the estimated values of the IPTW method display a strong variation with the frame orientation. A wide range of possibilities  is found for the LMC \omp,  from $0$ to $55\,$\kmskpc. Note also the clear correlation between the bar major and minor axis with the orientations where the shape of the \omp\ curve vary significantly. The median of all IPTW values seen in this graph is $23\pm 12\,$\kmskpc, adopting  here  the mean absolute deviation as the uncertainty.

\begin{figure}
    \centering
    \includegraphics[width=0.5\textwidth,clip, trim=0cm 1.25cm 0cm 0cm]{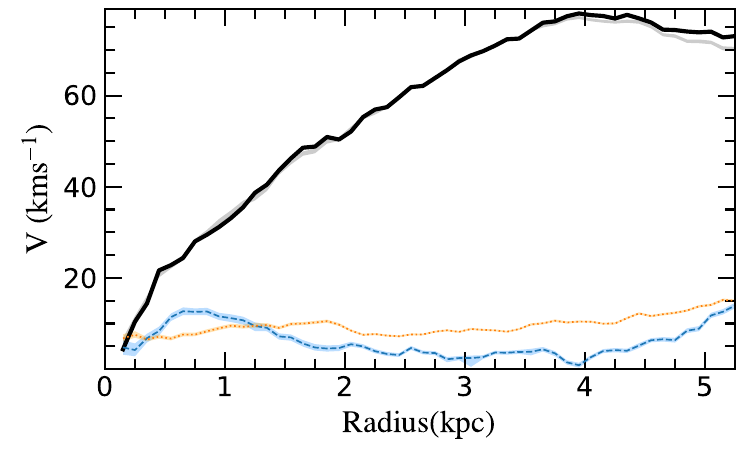}
    \includegraphics[width=0.5\textwidth,clip, trim=0cm 1.25cm 0cm 0cm]{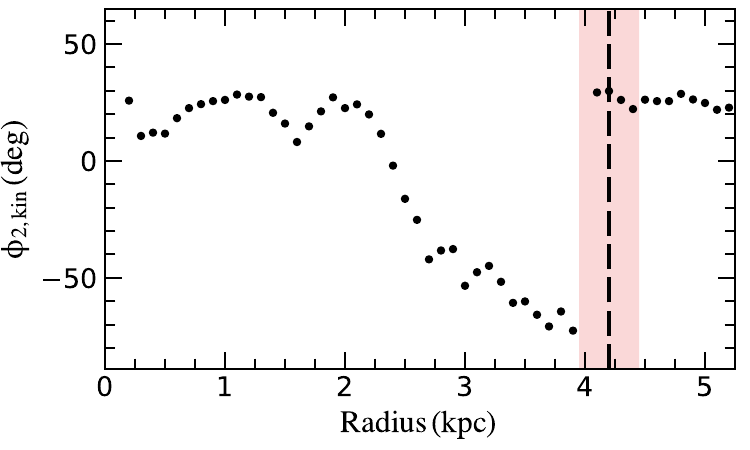}
    \includegraphics[width=0.5\textwidth]{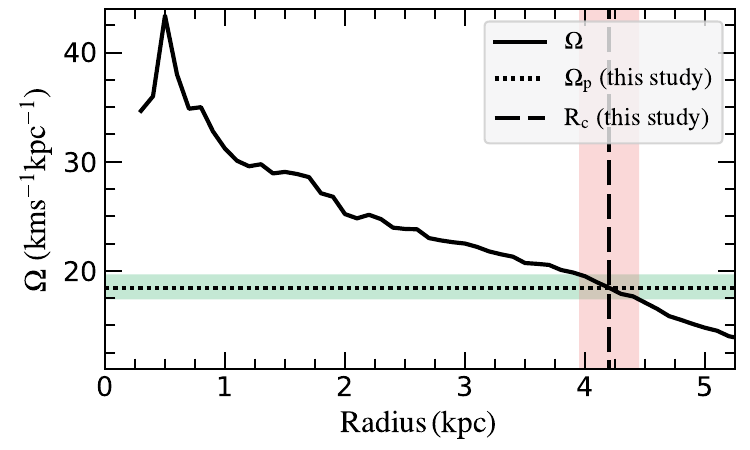}
    \caption{Results of the BV model of the stellar tangential velocity map of the LMC NN complete sample: strength (upper row) and phase angle (lower row, $\phi_{2,\rm kin}$) of the bisymmetric Fourier mode. The black solid line is the fitted axisymmetric velocity component (the rotation curve), the grey line is the median velocity (initial value for the model), the blue dashed curve is the strength of the tangential bisymmetry, and the orange dotted line is the scatter of the model. The vertical light coral area and dashed line shows the adopted bar corotation radius of the LMC, $R_c = 4.20 \pm 0.25$ kpc. In the bottom panel, we show the angular velocity of the LMC as a function of radius. The vertical light coral area and dashed line show the corotation radius of the bar. The horizontal green area and dotted line shows the corresponding bar pattern speed $\Omega_p = 18.5_{-1.1}^{+1.2}$ \kmskpc. }
    \label{fig:fourier_lmc_corotation}
\end{figure}

Finally, in Fig.~\ref{fig:fourier_lmc_corotation} we show the results of the BV method applied to the LMC tangential velocity map of Fig.~\ref{fig:dehnen_lmc}.  The LMC rotation curve from the NN complete sample (upper panel,  light grey) is very similar to the 0th order Fourier component of the BV model (black solid line). The amplitude of the LMC bar perturbation is stronger at $R=0.75$ kpc (blue line). The orange dotted line showing the scatter in the residual tangential velocity is often larger than the bisymmetric mode. It thus shows that the bar is not the only perturber in the LMC disc, but this does not prevent the bisymmetry from being detected efficiently by the method. Note that seeing the scatter in the model stronger with radius  is reminiscent to the finding of the KRATOS simulation (top right panel of Fig.~\ref{fig:fourier_simu_corotation}). This is consistent with the observed complex stellar morphology in this region. A roughly constant value of $\phi_{2,\rm kin} \simeq 15-20\degr$ is seen out to $R = 2$ kpc, in good agreement with the phase angle of the bisymmetry of the density (Fig.~\ref{fig:dehnen_lmc_amplitude}, bottom panel), followed with a smooth decrease out to $R = 3.95$ kpc, as  evidence of the impact of  arms in the kinematics even in the bar region. This radius is the location from where the amplitude $V_2$ starts to increase.  At this radius, $\phi_{2,\rm kin}$ changes by $\sim 100\degr$ to recover a constant value comparable to the bar phase angle at low radius. Following prescriptions from the numerical modelling, we adopt the radius just after the sharp transition of phase angle as the bar corotation radius, placing the LMC bar corotation at $R_c = 4.20 \pm 0.25$ kpc.   Relative to the  angular velocity curve $\Omega$ (solid line in the bottom panel of Fig.~\ref{fig:fourier_lmc_corotation}), it corresponds to a LMC bar pattern speed of $\Omega_p = 18.5_{-1.1}^{+1.2}$ \kmskpc. 

\begingroup

\setlength{\tabcolsep}{10pt} % Default value: 6pt
\renewcommand{\arraystretch}{1.5} % Default value: 1
\begin{table}
\centering
\begin{tabular}{llll}
\hline
Method                &  $R_1$  & $R_c$ & $\Omega_p$ \\ \hline
LTW                   &             &          &  30.3 $\pm$ 1.3\\ 
IPTW                  &             &          &  23.1 $\pm$ 12\\ 
Dehnen                &  2.3   &          &  $-1.0 \pm 0.5$ \\ 
BV                   &             &    4.20 $\pm$ 0.25   &  $18.5^{+1.2}_{-1.1}$  \\ \hline
\end{tabular}
\caption{Results of applying the method used in this work to the LMC complete sample. Bar radius and bar corotation are in kpc and the bar pattern speed is in \kmskpc. }
\label{tabl:summary_LMC}
\end{table}

\endgroup

\section{Discussion}
\label{sec:discussion}

Table~\ref{tabl:summary_LMC} is a summary of the LMC bar properties  obtained  with the different methods in the previous section. At first glance, it is difficult to conclude the pattern speed of the stellar bar of the LMC given the large range of values. Only a few studies have provided estimates of the LMC bar pattern speed. \citet{Shimizu2012} derived a value for the bar pattern speed  based on the idea that the Shapley Constellation III star forming region \citep{Shapley1951} is located at the $L_4$ Lagrangian point  of the non-axisymmetric bar potential rotating frame. The authors found $\Omega_p=21\pm 3\,$\kmskpc. Unfortunately, they did not report on the distance to the LMC they have assumed, which makes the comparison with our results not trivial. Nevertheless, the value they quote is in good agreement with the one inferred from the BV method. In another work, \citet{Wan2020} used SkyMapper \citep{wolf18} data to study the internal kinematics of the LMC populations, following coordinate transformations  described in \citet{vandermarel01} and \citet{vandermarel02}, thus, the same transformations we applied here. For their Carbon Stars, \citet{Wan2020} fit a rotation curve built on a constant angular speed of stars as a function of radius. This may seem a simplistic assumption because $\Omega$ must vary with radius (see e.g. Fig.~\ref{fig:fourier_lmc_corotation}). Therefore, they did not constrain the pattern speed of the bar, but their result gives an  estimate of what the rough angular frequency of stars should be within $R \sim 8$ kpc, thus on \omp\ since at corotation $\Omega$ equals the desired pattern speed. They found $\Omega = 24.6\pm 0.6$ \kmskpc, which is not far from the BV value derived in our work.

A second important result is that  the TW method  is extremely sensitive to the orientation of the x-y frame, and therefore to the way the integrals view the bar perturbation in the disc.  Finding a dependency of the TW integrals with viewing angles in galaxies is not a new result. Using a numerical simulation, \citet{2019zou} found that the accuracy on bar pattern speeds could be kept under $\sim 10\%$ for a bar orientated by 10\degr-75\degr\ and 105\degr-170\degr\ with respect to the reference axis of the disc, while configurations where the integrals are measured perpendicularly to the bar major axis were shown to imply values systematically different from reality. This led for instance \citet{2019cuomo} to define their sample of barred galaxies with bar position angles by 10\degr\ or more apart from the disc major and minor axes. With our simulations, although frame orientations  near the principal axes of the bar should  be avoided, which agrees with the findings of \citet{2019zou}, the results of Sect.~\ref{sec:simulations} and~\ref{sec:results} did not allow us to identify any particular wide range of orientations where the bar pattern speed estimates are reliable, and that the probability of agreement is low. It is worth mentioning here that bar pattern speeds of galaxies, as  measured with the LTW method, are also known to be sensitive to the orientation of the pseudo-slits with respect to the disc line-of-nodes \citep{2003debattista}. Still with the help of numerical simulations, these authors showed  that assuming an incorrect position angle for the disc major axis  can lead to large errors on \omp\ when performing the numerical LTW integrals. Of course, this is not directly linked to the viewing angle of the bar itself in the considered disc plane, as shown above in the IPTW case, but it illustrates nicely how sensitive to orientations the TW method can be.

The origin of the strong variations with the bar angle, and of the large discrepancy with true values, may be the impact of patterns other than the bar in the TW integrals, like   spiral arms in the N-body simulation and the LMC. A possible solution to overcome this issue could be to measure \omp\ as a function of radius, as done in \citet{2006merrifield} or \citet{2008meidt-b} for other galaxies. However, such analysis is beyond the scope of this article.  

\begin{figure}
    \centering
    \includegraphics[width=\columnwidth]{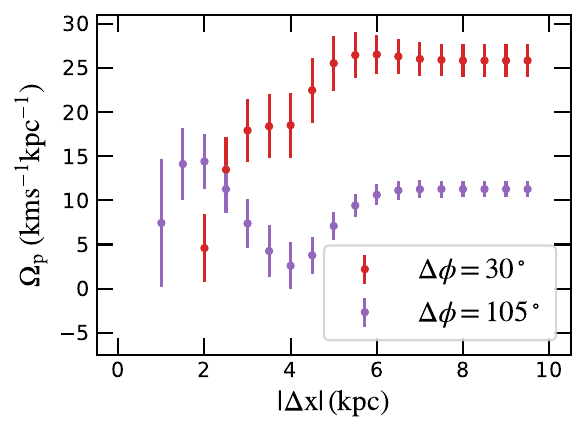}
    \caption{Convergence tests of the LMC pattern speed from the IPTW method. $|\Delta x|$ is the domain of the numerical integrations of the TW equations (Eq.~\ref{eq:tw_vxvy}). The convergence has been measured at two orientations of the frame $\Delta\phi=30\degr$ and $105\degr$. }
    \label{fig:tw_convergence}
\end{figure}

Nevertheless, the impact of other patterns than the bar on \omp\ can be tested by studying the convergence of the TW integrals as a function of the aperture $|\Delta x|$ in which the  integrals are measured. This is achieved by progressively increasing $|\Delta x|$ \citep[see e.g.][]{2009chemin, 2019zou}. In particular, if the outer LMC spiral arms contaminate the TW integrals when the maximum range of $|\Delta x|$ allowed by the extent of the observation is adopted (as we did in previous sections), then the derived pattern speed  is mixing both the bar and spiral patterns, and the bar speed may be underestimated. Indeed,  pattern speeds of spiral arms are expected to be lower than those of bars \citep{2006merrifield, 2019zou}. But with a smaller domain of integration, chosen wisely, the pattern speed could converge to another \omp, that of the bar only. We thus varied  $|\Delta x|$ within 1 to 9 kpc in the LMC, for two examples of frame orientations, $\Delta\phi =30\degr$ and  $\Delta\phi = 105\degr$ (directions outside the LMC principal bar axes, see Fig.~\ref{fig:tw_rot_LMC}). Figure~\ref{fig:tw_convergence} presents results of this test for the IPTW method. We find that  the integrals have converged at $|\Delta x| = 6$ kpc at the constant values reported in Fig.~\ref{fig:tw_rot_LMC} at the selected $\Delta\phi$   ($\sim 25$ and $\sim 10$ \kmskpc, respectively). But, for $|\Delta x| < 6$ kpc, \omp\ varies significantly, either increasing and/or decreasing. In other words, we do not find  hints of secondary convergence regions of the integrals that would correspond to the LMC bar \omp\ only. To the benefits of the TW method, we can nonetheless see that a rough LMC pattern speed \omp\ found by averaging the TW values over all $\Delta\phi$ orientations is 23.1 \kmskpc, but with a large scatter of 12 \kmskpc. This compares  with the value found with the BV model, but not with the one from the Dehnen method.

As consequence, without  agreement among the trends found with  the various simulations used in the previous works and our present study, and without an identifiable region of bar orientation where ground-truth and measured pattern speeds agree within simulations, the individual  pattern speeds found by the TW method in Fig.~\ref{fig:tw_rot_LMC} cannot be representative of the real LMC bar \omp.  
The agreement of \omp\ found by the LTW method with the value found by the IPTW method for the LMC also indicates that the pattern speed of bars measured by means of the LTW method may likely be only representative of any value stemming from random frame orientations fixed by the position angle of the major axis of discs on the sky plane, but not of a global bar angular frequency. 
The reason of the failure of the TW method in giving a coherent LMC bar pattern speed is unclear. It could be that the tidal interaction with the SMC has broken the conditions of applicability of the method, the disc being no more in full equilibrium. However, both isolated and interacting discs in the simulations show similarities with the observations. It could also indicate that the impact of the spiral arms on the TW integrals is not as negligible as initially thought. More work will be necessary to investigate  the origin of this issue with the TW method applied to the simulations and the LMC data.

The pattern speed obtained with the Dehnen method is significantly different from the one of the BV model. Unlike the TW method, we have shown that the Dehnen method  performs nicely with an idealised simulation with a well defined rotationally supported stellar disc and no external perturbations and a simulation of an interacting and out-of-equilibrium disc, as it is insensitive to the bar orientation and the outer spiral perturbation, by construction. Applied to the LMC NN complete sample, it surprisingly results in a bar with null rotation, perhaps slightly counter-rotating with respect to the LMC disc.  Is this finding realistic? Peculiar bars with such property exist in numerical simulations. In a recent  work, it was shown how a bar embedded in a counter-rotating dark matter halo can decelerate, then flip its pattern speed, and finally decoupling its rotation from the disc \citep{2023Collier}.  After the sign flip, the bar suffers from a large inclination and develops a warped disc. This scenario is difficult to test in real data because we need an observable to check the dark matter halo rotation. Evidence exists that LMC disc is warped \citep[e.g.][]{Choi2018,ripepi22}, however we do not observe in the kinematic maps a decoupling of the bar motion from that of the disc. Another possible origin of the bar deceleration and counter-rotation could be external and due to the interaction with the SMC and/or the MW. This scenario could be tested with appropriate numerical simulations such as the KRATOS suite of simulations of LMC-SMC-MW-like tidal encounters (Jim\'enez-Arranz et al., in preparation).

This result does not come without issues, however. An almost non-rotating LMC bar would indeed not show any corotation within the disc since such \omp\ should never cross the $\Omega$ curve. It is not an easy task to imagine how the orbits and the disc structure would respond to this peculiar circumstance. An absence of corotation could allow the bar to increase its length and strength out to the  disc outskirts, that is, make the orbits of stars and the LMC stellar gravitational potential very elongated throughout the whole LMC disc. Indeed, nothing could prevent it here from growing significantly owing to the absence of corotation and the expected destructive orbits perpendicular to the bar beyond corotation. The LMC stellar density map  shows that the outer LMC disc is elliptical \citep[see Fig.~\ref{fig:dehnen_lmc}, and also e.g.][]{luri20}, but the elongation occurs along a direction that is not aligned with the LMC bar. The elongated potential of the LMC likely comes from the tidal interaction with the MW and the SMC. Furthermore, the absence of bar resonances in the inner disc in this framework would make it difficult to interpret  the strong variation of the orientation of the velocity bisymmetry evidenced in Fig.~\ref{fig:fourier_lmc_corotation}, which is expected to occur naturally around corotation. We think that the method may be sensitive to dust extinction and completeness effects in the inner LMC region, perhaps more strongly than the other methods. Also, the inner disc is not fully traced by a bar pattern, and the density map clearly shows pieces of spiral arms inside the circle encompassing the bar region.  The inner kinematics is not fully dominated by the bar either, due to the smoothly varying phase angle of the velocity bisymmetry within $R=2-4$ kpc (see Fig.~\ref{fig:fourier_lmc_corotation}), likely caused by a winding spiral structure near the tips of the bar. All of these effects may hamper the method from yielding a  \omp\ representative of the bar.

Assuming that the corotation radius $R_c = 4.20 \pm 0.25$ kpc measured by the BV model is more representative of the bar properties, it corresponds to a pattern speed of $18.5_{-1.1}^{+1.2}$ \kmskpc. The LTW pattern speed of $30.4 \pm 1.3$ \kmskpc\ would thus be discrepant by 64\% from the one inferred here. When compared to its radius of 2.3 kpc, the LMC stellar bar has $R_c / R_1 = 1.8 \pm 0.1$, thus corresponding to a slow bar, according to numerical methods \citep{Athanassoula1992}. Finally, if we assume that the pattern speed has to be estimated using a velocity curve tracing more closely the circular velocity \citep[the rotation curve of the younger stellar populations in][]{jimenez-arranz23} than the tangential velocity of the whole sample dominated by older stars (upper panel of Fig.~\ref{fig:fourier_lmc_corotation}), then $R_c = 4.2$ kpc would translate into $\Omega_p = 20.9 \pm 1.1$ \kmskpc, which still compares well with the value found for the whole sample.

\section{Conclusions}
\label{sec:conclusions}

In this work we use three different methods to determine the LMC bar pattern speed, namely the TW method in its original form, when only line-of-sight velocities are available (LTW), and a variation of it that also makes use of astrometric data and in-plane velocity fields (IPTW); the Dehnen method, which is recently published and tested using single snapshots of N-body simulations; and the bisymmetric velocity (BV) method, which is based on the Fourier decomposition of the tangential velocity of a bisymmetric model to constrain the corotation radius of the bar. In order to characterise the strengths and limitations of each of the methods, we applied them to two different simulated barred galaxies. One snapshot of an N-body simulation of an isolated disc galaxy \citep[B5,][]{rocafabrega2013} and one snapshot of a N-body simulation of an interacting disc galaxy (KRATOS, Jiménez-Arranz et al., in preparation). The results show that: 

\begin{itemize}
    \item The TW method shows a large dependency on the frame orientation when applied to both B5 and KRATOS simulations. 
    \item The Dehnen method recovers with good accuracy and precision the true pattern speed when applied to both: an idealised simulation with a well defined rotationally supported stellar disc and no external perturbations and a simulation of an interacting and out-of-equilibrium disc.
    \item The BV method determines the corotation radius and pattern speed of both simulations. The accurate constraint of the strong variation of the kinematic phase angle  is crucial for the determination of the corotation radius and thus the bar pattern speed.
\end{itemize}

From these points, and when applying the methods to the LMC sample, we are inclined to:
\begin{itemize}
    \item Discard the pattern speeds found with  the TW method, because no obvious privileged value is found with the IPTW method, owing to the strong variation of the  integrals with the orientation of the $x-y$  plane, thus with the bar viewing angle inside the LMC disc. Also, the unique pattern speed found by the LTW method cannot be representative of a global bar frequency either. 
    \item Evaluate the validity of the bar pattern speed obtained with the Dehnen method. It corresponds to a non-rotating bar, with implications hard to reconcile with the structure and kinematics of the LMC disc.
    \item Provide a first tentative value of the LMC bar corotation radius at $R=4.20\pm 0.25$ kpc with the BV method, as the sharp change of the kinematic phase angle  measured through a Fourier modelling is very reminiscent to the signature of corotation seen in numerical simulations. It gives a  bar corotation-to-size ratio of  $R_c / R_1 = 1.8 \pm 0.1$, which places the LMC bar in the slow rotation regime. The corresponding LMC bar pattern speed is $\Omega_p = 18.5_{-1.1}^{+1.2}$ \kmskpc, reasonably consistent with other estimates found in the literature.
\end{itemize}

This research has presented novel constraints on the corotation and  pattern speed of the stellar bar of the LMC. Our intention is to continue this investigation, taking advantage of forthcoming releases from the \gaia mission that will offer improved data quality. With enhanced angular resolution, more precise proper motion measurements, and increased access to line-of-sight velocities, we anticipate  it will become easier to establish an LMC sample  with reduced limitations,  such as the crowding  of stars  at low radius, or contamination from foreground MW stars. Moreover,  working with the 3D velocities of LMC stars will offer new opportunities. These approaches have already been initiated by \citet{jimenez-arranz23}. By doing so, we will obtain more reliable estimates of the LMC pattern speed and potentially alleviate tensions that exist among the Dehnen and bisymmetric velocity methodologies tested in this study, which may arise because of the perturbed equilibrium of the LMC. This could be tested with various numerical simulations of the LMC, SMC and MW  encounters.
We will also study possible variations of the bar pattern speed among various stellar evolutionary phases of the LMC. 

\section*{Acknowledgements}
We are grateful to W. Dehnen, M. Semczuk and V. P. Debattista for fruitful discussions. We thank Dr A.G.A. Brown for his comments on the completeness estimation. We thank an anonymous referee for a critical review and constructive suggestions that helped improving the manuscript. This work has made use of data from the European Space Agency (ESA) mission {\it Gaia} (\url{https://www.cosmos.esa.int/gaia}), processed by the {\it Gaia} Data Processing and Analysis Consortium (DPAC, \url{https://www.cosmos.esa.int/web/gaia/dpac/consortium}). Funding for the DPAC has been provided by national institutions, in particular the institutions participating in the {\it Gaia} Multilateral Agreement.
OJA acknowledges funding by l'Agència de Gestió d'Ajuts Universitaris i de Recerca (AGAUR) official doctoral program for the development of a R+D+i project under the FI-SDUR grant (2020 FISDU 00011).
OJA, MRG and XL acknowledge funding by the Spanish MICIN/AEI/10.13039/501100011033 and by "ERDF A way of making Europe" by the “European Union” through grant RTI2018-095076-B-C21, and the Institute of Cosmos Sciences University of Barcelona (ICCUB, Unidad de Excelencia ’Mar{\'\i}a de Maeztu’) through grant CEX2019-000918-M.
LC acknowledges financial support from the Chilean Agencia Nacional de Investigaci\'{o}n y Desarrollo (ANID) through the Fondo Nacional de Desarrollo Cient\'{\i}fico y Tecnol\'{o}gico (FONDECYT) Regular Project 1210992 and the Institute of Cosmos Sciences University of Barcelona (ICCUB, Unidad de Excelencia ’Mar{\'\i}a de Maeztu’) through grant CEX2019-000918-M which founded a two-weeks visit in Barcelona.
PA and LC warmly acknowledge the financial support of the Comit\'e Mixto European Southern Observatory-Gobierno de Chile.
SRF acknowledges support from the Knut and Alice Wallenberg Foundation and the Swedish Research Council (grant 2019-04659).
PM gratefully acknowledges support from project grants from the Swedish Research Council (Vetenskapr\aa det, Reg: 2017-03721; 2021-04153).

\bibliographystyle{aa}
\bibliography{mylmcbib} % if your bibtex file is called example.bib

\begin{appendix}

\section{Test-particle simulation}
\label{sec:app-tpsim}

In this section, we apply the methods described in Section~\ref{sec:methods} to an idealised simulation of a barred disc galaxy in statistical equilibrium with the imposed potential. We briefly describe the characteristics of the simulation and the results of the methods applied to it.

We use a $5$ million test particle simulation with initial conditions, galactic potential, and steps performed in the integration process as described in \citet{RomeroGomez2015}. We refer to this simulation as the TP simulation. The initial conditions for positions and velocities were drawn for a disc density distribution following a Miyamoto-Nagai disc potential \citep{Miyamoto1975} with a typical scale-height ($h_z=300$\,pc) and radial velocity dispersion ($\sigma_{U}=30.3\,$\kms) of a red clump star. Then we integrate the initial conditions in the axisymmetric potential of \citet{Allen1991} for $10\,$Gyr, after that we introduce the Galactic bar potential adiabatically during $\text{four}$ bar rotations, and we integrated another $\text{16}$ bar rotations so that the particles achieve a statistical equilibrium with the final bar potential. The galactic bar consists of the superposition of two aligned Ferrers ellipsoids \citep{Ferrers1877}, one modelling a triaxial bulge with a semi-major axis of $3.13\,$kpc, and the second modelling a long thin bar with a semi-major axis of $4.5\,$kpc, with an angular orientation of $20^{\circ}$. In the TP simulation, we impose that the bar rotates counter-clockwise as a rigid body with a constant pattern speed of $42$ \kmskpc, placing the corotation resonance at $R_c = 4.9\,$kpc, measured as in the B5 simulation. The TP simulation represents the ideal configuration to estimate a bar pattern speed with the different methods due to the only barred perturbation and being in statistical equilibrium with the imposed potential.

\begin{figure*}
    \centering
    \includegraphics[width=1\textwidth]{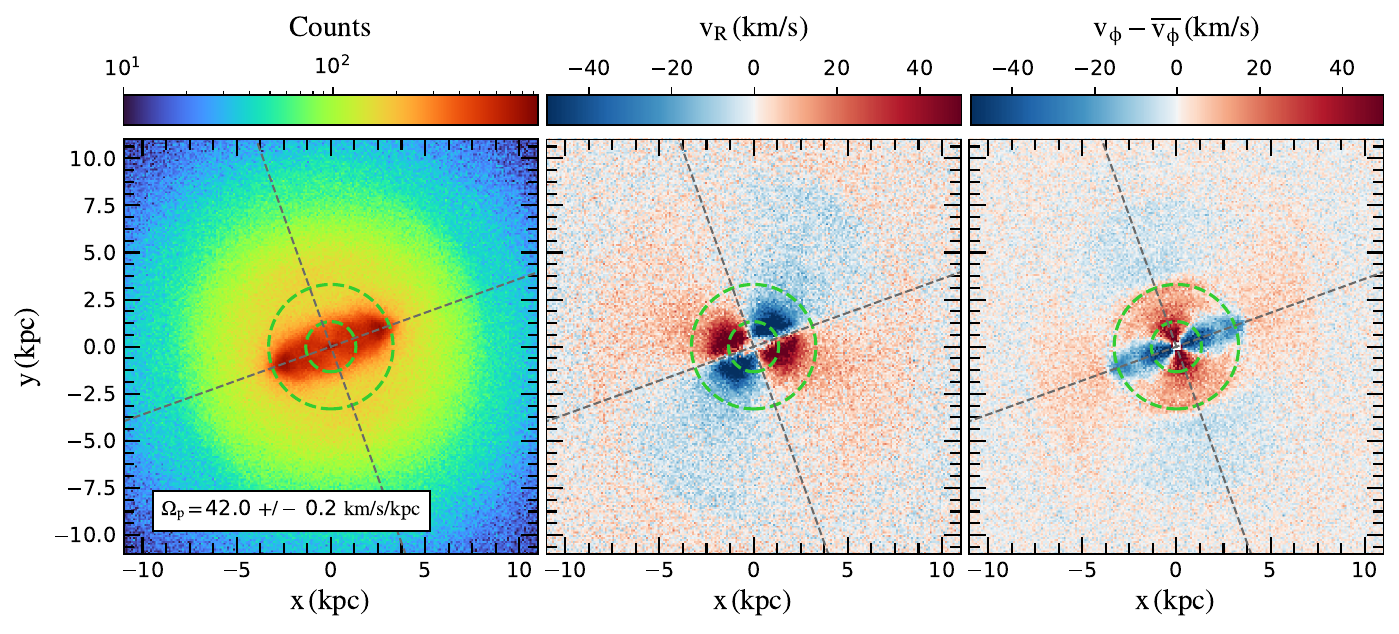}
    \caption{Application of the Dehnen method to the TP simulation. Surface density (left), median radial velocity map (center) and median residual tangential velocity map (right). The bar region identified by Dehnen method (see values in Table~\ref{tabl:dehnen_TP}) is indicated by green dashed circles. The grey dashed lines trace the bar minor and the major axes.}
    \label{fig:dehnen_TP}
\end{figure*}

Figure~\ref{fig:dehnen_TP} shows the surface density (left panel), the radial velocity (middle panel) and the residual tangential velocity (right panel) maps for the TP simulation. The map of the residuals has been obtained by subtracting the rotation curve to the \vp\ map. We observe the bar and no obvious spiral arms, which is expected because only a bar potential is modelled. In both radial and residual tangential velocity maps, we also  observe a kinematic quadrupole caused by the stellar orbits shaping the bar. In every panel, we highlight the bar region identified by the Dehnen method by green dashed circles, with  inner and outer circles corresponding to $R_0$ and $R_1$, respectively. The grey dashed lines trace the bar minor and the major axes found by the method. The bar orientation $\phi_b$ is in agreement with the orientation observed in the surface density, and separates remarkably  the quadrupole patterns in two parts. The Dehnen method infers a value of $\Omega_p = 42.0 \pm 0.2$ \kmskpc, in  agreement with the imposed value. Values are summarised in Table~\ref{tabl:dehnen_TP}.

\begingroup

\setlength{\tabcolsep}{10pt} % Default value: 6pt
\renewcommand{\arraystretch}{1.5} % Default value: 1
\begin{table*}
\centering
\begin{tabular}{llllllll}
\hline \hline
% \toprule
\multicolumn{1}{c}{} & \multicolumn{1}{c}{Reference} & \multicolumn{4}{c}{Dehnen method} & \multicolumn{2}{c}{BV method} \\
\cmidrule(rl){2-2} \cmidrule(rl){3-6} \cmidrule(rl){7-8}
     % & Reference &   &  &  &  \\
% \hline
 Simulation               & $\Omega_p$ &  $R_0$  & $R_1$ & $\Omega_p$ & $\phi_b$ & $R_c$ & $\Omega_p$ \\ \hline
TP  & 42.0 & 1.33 & 3.30 & 42.0 $\pm$ 0.2 & 19.8 $\pm$ 0.1 & 4.8 $\pm$ 0.5 & 43.3$^{5.0}_{-4.2}$ \\
\hline \hline
% \bottomrule
\end{tabular}
\caption{Results of the Dehnen and the BV methods applied to TP simulation, compared to the reference value (obtained using finite-differences). The inner, outer and corotation radii $R_0$, $R_1$ and $R_c$ are in kpc. The bar pattern speed \omp\ and phase angle $\phi_b$ are in \kmskpc and degrees, respectively.}
\label{tabl:dehnen_TP}
\end{table*}

\endgroup

\begin{figure}
    \centering
    \includegraphics[width=\columnwidth]{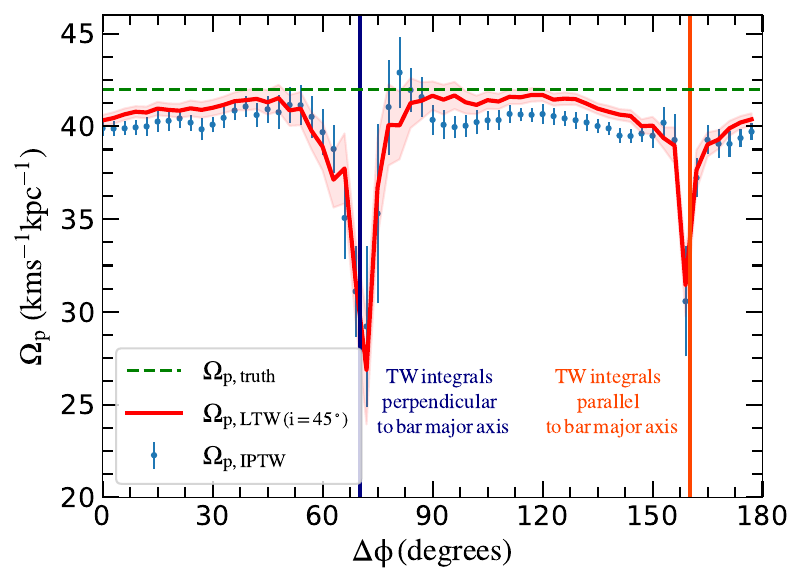}
    \caption{Same as in Fig.~\ref{fig:ltw_iptw} but for the TP simulation.}
    \label{fig:ltw_iptw_TP}
\end{figure}

Figure~\ref{fig:ltw_iptw_TP} shows the impact of the variation of the orientation of the Cartesian frame on the derived pattern speed using the LTW and IPTW methods, by rotating the reference $x$ and $y$ axes  around the $z-$axis in the simulation. The rotation of the Cartesian frame before sky projection allows the TW integrals (measured parallel to the major axis) to view the bar and spiral perturbations through various angles. Note that in this case, where no other non-axisymmetric component but the bar, and being in statistical equilibrium, the recovered pattern speed using both versions of the TW method show comparable trends. As a known issue of the TW, the integrals do not converge when the slit is aligned with the bar axes \citep{tw84}. The IPTW shows a  pattern speed systematically lower by $\sim 10\%$ from the true value, about twice as large as the systematic shown by the LTW method.

\begin{figure*}
    \centering
    \includegraphics[width=0.8\textwidth]{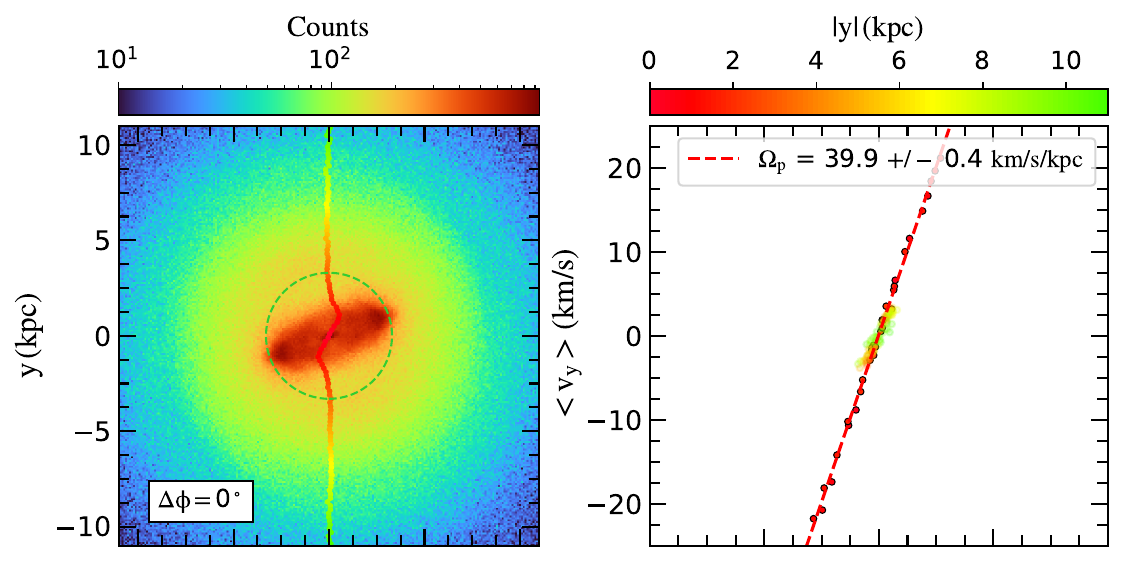}
    \includegraphics[width=0.8\textwidth]{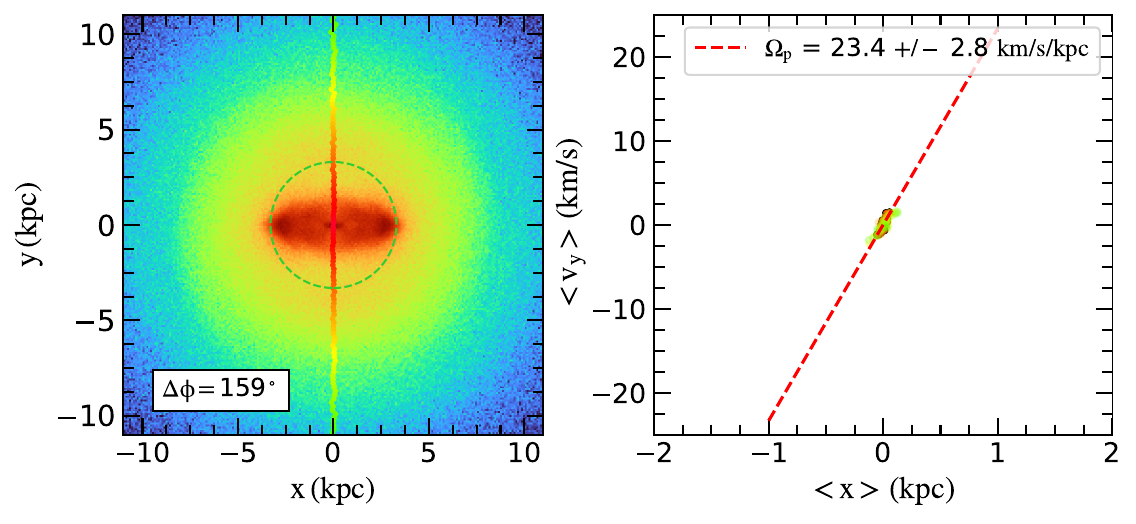}
    \caption{Same as in Fig.~\ref{fig:tw_B5} for the TP simulation. }
    \label{fig:tw_TP}
\end{figure*}

Figure~\ref{fig:tw_TP} presents results of the IPTW method for the TP simulation. Again, we show here two different reference frame orientations: the original one, with $\Delta\phi=0$\degr (top row) and the one with the bar major axis parallel to the $x$-axis, $\Delta\phi=159$\degr (bottom row), for illustrative purposes. In the first case, there is a clear linear trend and the IPTW recovers a bar pattern speed $\Omega_p = 39.9 \pm 0.4$ \kmskpc, with a relative difference smaller than $5\%$ from the imposed value. In the second case, the TW $x-$integrals are perfectly aligned with the bar major axis, as expected in the presence of only a bar potential. The $\langle x \rangle$ values are then very close to zero, so there is not a clear trend in this case (it would give similar results when the TW integrals are evaluated at another viewing angle along the bar minor axis). We recover here a counter-clockwise pattern speed of $\Omega_p = 23.4 \pm 2.8$ \kmskpc\ which differs by almost $50\%$ from the imposed value. The TW method thus performs better when no prominent sub-structures exist in the disc, as a well-known issue, integrals should not be made along the major or minor axes of the bar.

\begin{figure}
    \centering
    \includegraphics[width=0.48\textwidth]{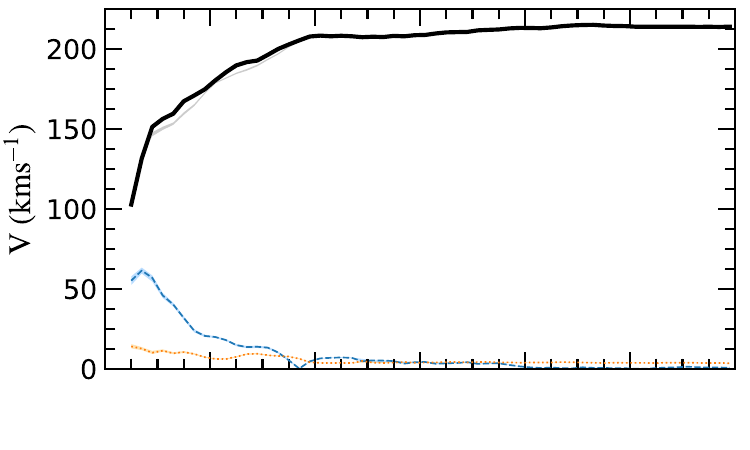}\\
    \vspace{-0.75cm}
    \includegraphics[width=0.48\textwidth]{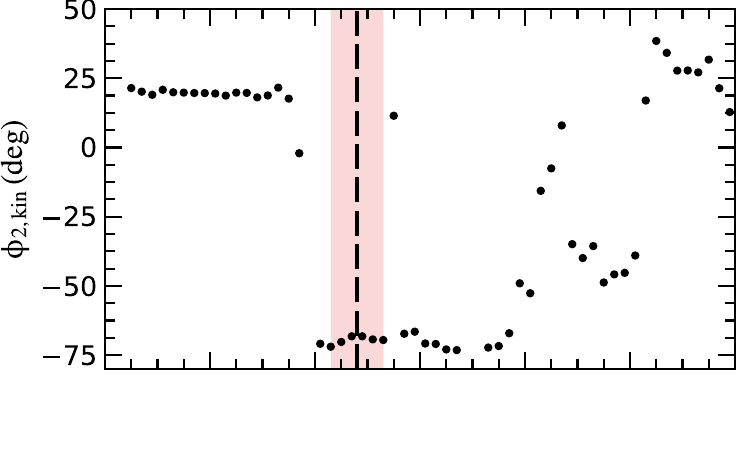}\\    
    \vspace{-0.75cm}
    \includegraphics[width=0.48\textwidth]{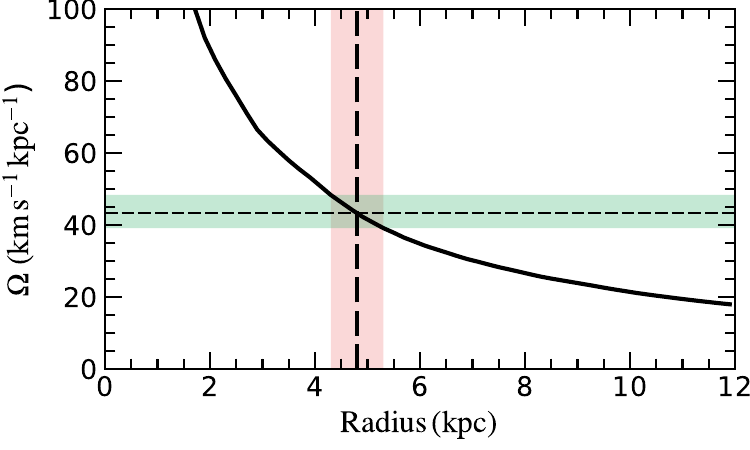}
    \caption{Same as in Fig.~\ref{fig:fourier_simu_corotation} but for the TP simulation. }
    \label{fig:fourier_TP_corotation}
\end{figure}

Finally, in Fig.~\ref{fig:fourier_TP_corotation}, we show the results of applying the BV method to the TP simulation, the value for corotation we derive is $R_c = 4.8 \pm 0.5\,$kpc, and the bar pattern speed is $\Omega_p = 43.3_{-4.2}^{+5.0}$ \kmskpc, which exceeds the true value of 42 \kmskpc, although remaining comparable with it given the lower quoted uncertainties. Values are summarised in Table~\ref{tabl:dehnen_TP}.

\end{appendix}

% Don't change these lines
% \bsp	% typesetting comment
\label{lastpage}

\end{document}